\documentclass[]{agujournal2019}

\usepackage{url} %this package should fix any errors with URLs in refs.
\usepackage[finalnew]{trackchanges} %for better track changes. finalnew option will compile document with changes incorporated.
\usepackage{soul}
\usepackage{gensymb}
\usepackage{amsmath}
\addeditor{R1}
\addeditor{R2}
\addeditor{mazur}
\addeditor{petr}
\addeditor{PBrown}
\addeditor{rob}
\addeditor{denis}
\addeditor{carsten}
\addeditor{gunter}
\addeditor{anamika}

\draftfalse

\journalname{Radio Science}

\begin{document}
\newcommand\aj{AJ}
          % Astronomical Journal
\newcommand\actaa{Acta Astron.}
  % Acta Astronomica
\newcommand\araa{ARA\&A}
          % Annual Review of Astron and Astrophys
\newcommand\apj{ApJ}
          % Astrophysical Journal
\newcommand\apjl{ApJ}
          % Astrophysical Journal, Letters
\newcommand\apjs{ApJS}
          % Astrophysical Journal, Supplement
\newcommand\ao{Appl.~Opt.}
          % Applied Optics
\newcommand\apss{Ap\&SS}
          % Astrophysics and Space Science
\newcommand\aap{A\&A}
          % Astronomy and Astrophysics
\newcommand\aapr{A\&A~Rev.}
          % Astronomy and Astrophysics Reviews
\newcommand\aaps{A\&AS}
          % Astronomy and Astrophysics, Supplement
\newcommand\azh{AZh}
          % Astronomicheskii Zhurnal
\newcommand\baas{BAAS}
          % Bulletin of the AAS
\newcommand\caa{Chinese Astron. Astrophys.}
  % Chinese Astronomy and Astrophysics
\newcommand\cjaa{Chinese J. Astron. Astrophys.}
  % Chinese Journal of Astronomy and Astrophysics
\newcommand\icarus{Icarus}
  % Icarus
\newcommand\jcap{J. Cosmology Astropart. Phys.}
  % Journal of Cosmology and Astroparticle Physics
\newcommand\jrasc{JRASC}
          % Journal of the RAS of Canada
\newcommand\memras{MmRAS}
          % Memoirs of the RAS
\newcommand\mnras{MNRAS}
          % Monthly Notices of the RAS
\newcommand\na{New A}
  % New Astronomy
\newcommand\nar{New A Rev.}
  % New Astronomy Review
\newcommand\pra{Phys.~Rev.~A}
          % Physical Review A: General Physics
\newcommand\prb{Phys.~Rev.~B}
          % Physical Review B: Solid State
\newcommand\prc{Phys.~Rev.~C}
          % Physical Review C
\newcommand\prd{Phys.~Rev.~D}
          % Physical Review D
\newcommand\pre{Phys.~Rev.~E}
          % Physical Review E
\newcommand\prl{Phys.~Rev.~Lett.}
          % Physical Review Letters
\newcommand\pasa{PASA}
  % Publications of the Astron. Soc. of Australia
\newcommand\pasp{PASP}
          % Publications of the ASP
\newcommand\pasj{PASJ}
          % Publications of the ASJ
\newcommand\qjras{QJRAS}
          % Quarterly Journal of the RAS
\newcommand\rmxaa{Rev. Mexicana Astron. Astrofis.}
  % Revista Mexicana de Astronomia y Astrofisica
\newcommand\skytel{S\&T}
          % Sky and Telescope
\newcommand\solphys{Sol.~Phys.}
          % Solar Physics
\newcommand\sovast{Soviet~Ast.}
          % Soviet Astronomy
\newcommand\ssr{Space~Sci.~Rev.}
          % Space Science Reviews
\newcommand\zap{ZAp}
          % Zeitschrift fuer Astrophysik
\newcommand\nat{Nature}
          % Nature
\newcommand\iaucirc{IAU~Circ.}
          % IAU Cirulars
\newcommand\aplett{Astrophys.~Lett.}
          % Astrophysics Letters and Communications
\newcommand\apspr{Astrophys.~Space~Phys.~Res.}
          % Astrophysics Space Physics Research
\newcommand\bain{Bull.~Astron.~Inst.~Netherlands}
          % Bulletin Astronomical Institute of the Netherlands
\newcommand\fcp{Fund.~Cosmic~Phys.}
          % Fundamental Cosmic Physics
\newcommand\gca{Geochim.~Cosmochim.~Acta}
          % Geochimica Cosmochimica Acta
\newcommand\grl{Geophys.~Res.~Lett.}
          % Geophysics Research Letters
\newcommand\jcp{J.~Chem.~Phys.}
          % Journal of Chemical Physics
\newcommand\jgr{J.~Geophys.~Res.}
          % Journal of Geophysical Research
\newcommand\jqsrt{J.~Quant.~Spec.~Radiat.~Transf.}
          % Journal of Quantitiative Spectroscopy and Radiative Trasfer
\newcommand\memsai{Mem.~Soc.~Astron.~Italiana}
          % Mem. Societa Astronomica Italiana
\newcommand\nphysa{Nucl.~Phys.~A}
          % Nuclear Physics A
\newcommand\physrep{Phys.~Rep.}
          % Physics Reports
\newcommand\physscr{Phys.~Scr}
          % Physica Scripta
\newcommand\planss{Planet.~Space~Sci.}
          % Planetary Space Science
\newcommand\procspie{Proc.~SPIE}
          % Proceedings of the SPIE
\let\astap=\aap
\let\apjlett=\apjl
\let\apjsupp=\apjs
\let\applopt=\ao
% TITLE

%%TC:ignore

\title{Precision measurements of radar transverse scattering speeds from meteor phase characteristics}

% AUTHORS

\authors{Michael Mazur\affil{1},
Petr Pokorn\'{y}\affil{2,3},
Peter Brown\affil{4,5},
Robert J. Weryk\affil{8},
Denis Vida\affil{1},
Carsten Schult\affil{6},
Gunter Stober\affil{9},
Anamika Agrawal\affil{7}}

\affiliation{1}{Department of Earth Sciences, University of Western Ontario, London, Ontario, N6A 3K7, Canada}
\affiliation{2}{Astrophysics Science Division, NASA/Goddard Space Flight Center, Greenbelt, Maryland, 20071, USA}
\affiliation{3}{Department of Physics and Astronomy, Catholic University of America, Washington D.C, 20064, USA}
\affiliation{4}{Department of Physics and Astronomy, University of Western Ontario, London, Ontario, N6A 3K7, Canada}
\affiliation{5}{Centre for Planetary Science and Exploration, University of Western Ontario, London, Ontario, Canada N6A 5B7}
\affiliation{6}{Leibniz Institute of Atmospheric Physics at the Rostock University, Schloss-Str. 6, 18225 K\"uhlungsborn, Germany}
\affiliation{7}{Department of Physics, University of California at San Diego,  9500 Gilman Dr. La Jolla, CA 92093, USA}
\affiliation{8}{Institute for Astronomy, University of Hawaii, 2680 Woodlawn Drive, Honolulu, HI, 96822, USA}
\affiliation{9}{Institute of Applied Physics, University of Bern, 3012 Bern, Switzerland}

\correspondingauthor{M. J. Mazur}{mmazur5@uwo.ca}

% KEYPOINTS

\begin{keypoints}
\item Meteor speeds \add[PBrown]{measured by radar}
\item Our algorithm allows determination of the meteor speed with 5\% uncertainty for $>90\%$ echoes using only a single radar station
\item We tested our method for two different radar systems with more than 10 million unique events
\item The entire code base is publicly available and well\add[R1]{-}documented

\end{keypoints}

%%TC:endignore

% ABSTRACT

\begin{abstract}
    We describe an improved technique for using the \change[mazur]{reflected phase from backscatter}{backscattered phase from} meteor radar echo measurements just prior to the specular point ($t_{0}$)\remove[R2]{,} to calculate meteor speeds and their uncertainty. Our method, which builds on earlier work of \citeA{Cervera_etal_1997}, scans possible speeds in the Fresnel distance \change[R1]{/}{-} time domain with a dynamic, sliding window and derives a best-speed estimate from the resultant speed distribution. We test the performance of our method, called pre-$t_{0}$ speeds by sliding-slopes technique (PSSST), on transverse scattered meteor echoes observed by the Middle Atmosphere Alomar Radar System (MAARSY) and the Canadian Meteor Orbit Radar (CMOR), and compare the results to time-of-flight and Fresnel transform speed estimates. Our novel technique is shown to produce good results when compared to both model and  speed measurements using other techniques. We show that our speed precision is $\pm$5$\%$ at speeds less than 40 km/s and we find that more than 90$\%$ of all CMOR multi-station echoes have PSSST solutions. For CMOR data, PSSST is robust against \change[R1]{$t_0$ point pick errors}{the selection of critical phase value} and poor phase unwrapping. Pick errors of up to $\pm$6 pulses for meteor speeds less than about 50\add[R2]{ }km/s produce errors of less than $\pm$5$\%$ of the meteoroid speed. In addition, the width of the PSSST speed Kernel density estimate (KDE) is used as a natural measure of uncertainty that captures both noise and $t_0$ pick uncertainties.
\end{abstract}

\section{Introduction}
\label{sec:intro}

Measurement of the velocity of meteoroids in Earth's atmosphere is of fundamental importance \add[R1]{to our understanding of solar system dynamics of small interplanetary bodies}. The velocity distribution of meteoroids is a major constraint for models of solar system dust production \cite<e.g.,>[]{Nesvorny_etal_2010,Grun_etal_1985}, is key to estimating dust input into the atmosphere \cite{Carrillo-sanchez_etal_2015,Carrillo-sanchez_etal_2016} and is crucial to models assessing the risk to spacecraft from meteoroid impacts \cite{Mcnamara_etal_2004}. Meteor speeds in Earth's atmosphere can be directly translated into equivalent orbits around the Sun \cite{Ceplecha_1987}\add[R1]{,} providing astronomical context to atmospheric meteor measurements. \add[R1]{As such, accurate measurement of meteor velocities is essential for determination of high-quality orbital elements that can be used to understand dynamics and evolution of small bodies in the Solar System.}

Optical systems rely on multi-station measurements of meteors to determine speeds for mm-sized particles and larger, while radar systems can measure speeds for particles approaching the ablation limit (sizes $\sim$ tens of microns) \cite{Ceplecha_1998}. Meteor speeds measured by radar have the added advantage of larger numbers than optical systems (radar systems typically detecting thousands of echoes per day) and \add[PBrown]{that} multiple independent speed estimation techniques \add[Gunter]{are available}. The main disadvantage of radar meteor speed measurements are the many biases which affect detection and hence make \change[R1]{absolute}{high-precision} estimates of radar speed distribution challenging \cite{Moorhead_etal_2017}.

Since the earliest observations of radar meteor echoes in the 1940's, a number of different techniques for determining radar meteor speeds have been developed. These early methods fall into two broad categories: those that relied on the interpretation of either radial scattering from meteor head echoes \cite{Hey_Stewart_1947,Hey_etal_1947}, or those using Fresnel diffraction \cite{Davies_Ellyett_1949} applied to transverse scattering from meteor trails \cite{Baggaley_2002}.

For radial scattering from head echoes, the radar wave is backscattered directly off the ionized region close (of order meters) to the meteoroid itself. This results in the meteor being observed as a moving target whose range-time relationship is hyperbolic. Head echoes have radar cross sections typically 60 dB smaller than transverse echoes \cite{Baggaley2009}, so this technique is only useful for detecting comparatively large meteoroids or through the use of high-power large-aperture radar systems, such as Milestone Hill, Arecibo, EISCAT, or MAARSY \cite{Evans1965, Mathews1997, Kero2008, Janches_Revelle_2005,SCHULT2017}.

For transverse scattering, the \add[PBrown]{speed measurement} technique with the longest heritage is the Fresnel amplitude diffraction method \cite{ELLYETT1948}. This approach uses the post-$t_{0}$ \add[R1]{(where $t_{0}$ refers to the echo time corresponding to the position along the trail when the moving meteoroid reaches the minimum range to the station, or specular point, as described below)} Fresnel oscillations in radar amplitude to compute meteor speed. Although it is useful for much smaller meteoroids\add[R2]{ than would be possible to detect with the same radar system measuring head echo scattering}, it measures only the most well-behaved (i.e. non-fragmenting) meteor echoes. A more recent spectral implementation, which is a hybridization of this technique, is described by \citeA{Hocking_2000}, who showed that the technique is able to measure speeds for $\approx$ 5\% of all echoes. These findings were confirmed by \citeA{Stober:2013} by comparing specular meteor speeds measured using the \citeA{Hocking_2000} approach to meteor head echo measurements. This underscores the primary limitation of the Fresnel amplitude speed technique - namely that fragmenting meteoroids result in the majority ($\sim 90\%$) of all echoes having amplitude oscillations too blurred to yield speeds \cite{Baggaley_2002}.  In contrast, \citeA{Baggaley_etal_1997} demonstrated that the amplitude-rise-time of specular echoes, which reflect the time taken to cross the first Fresnel zone, can yield speeds for virtually all echoes, albeit with comparatively low precision.

The most recently described single-station speed measurement technique for transverse scattering echoes is the Fresnel Transform (FT) method \cite{Elford_2004}. This approach is capable of measuring precise speeds and the one-dimensional distribution of scattering centers responsible for the time-amplitude profile of the echo, producing a "snapshot" of the ionization of the trail as a function of distance from the meteor head. \citeA{Holdsworth_etal_2004} demonstrates an automated approach to generating FT speeds.

\citeA{Cervera_1996} introduced a new technique for determining meteor speeds based on pre-$t_{0}$ phase information. While the method is inherently based on Fresnel diffraction, it relies on pre-$t_{0}$ phases and is not affected by fragmentation that tends to mask post-$t_{0}$ oscillations. As a result, it is usable for speed determination on about 75\% of echoes instead of about 10\% for the amplitude-based method \cite{Cervera_etal_1997}. By capitalizing on the fact that pre-$t_{0}$ phase is measurable even when the amplitude is at background levels, the method of \citeA{Cervera_etal_1997} provides a powerful tool for measuring speeds of single-station specular echoes. However, while \citeA{Cervera_etal_1997} provide details of the basic algorithm for pre-$t_{0}$ measurement, we find that in practice implementation of a pre-$t_{0}$ speed estimator is challenging as many details of an automated technique are not provided in the \add[R1]{published }literature. In particular, the algorithm requires identification of the timing of the $t_{0}$ point to get precise speeds  - this is unaddressed in these works. Indeed, in general no entirely reliable means of consistently finding $t_{0}$ is available, so a reproducible approach is required that estimates $t_{0}$ and its possible error and provides an associated speed uncertainty.

Starting in the late 1950's, the advent of computers led to the development of multi-station velocity solutions using forward scatter radar \cite{Gill_Davies_1956}. Although the early work of Gill and Davies required velocities to first be calculated using the diffraction technique to define common fiducial points on the echo profile, the later use of interferometry at the receivers \cite{Baggaley_etal_1994} meant that speeds could be calculated for a much larger number of meteors than was possible with the diffraction method. Extension of this technique more recently uses methods largely independent of the detailed amplitude or phase behaviour of the meteor echo. This is the geometrical time-of-flight (TOF) solution generated by modern-day meteor orbit radar systems such as the Canadian Meteor Orbit Radar (CMOR) \cite{Webster_etal_2004,Jones_etal_2005} or the Southern Argentina Agile Meteor Radar (SAAMER) \cite{Janches_etal_2015}. These require only a common timing fiducial point on multi-station measurements of the same echo (such as the amplitude inflection point) to generate full velocity vectors \cite{Baggaley_etal_1994}. Details of this technique as implemented with CMOR are provided in Appendix A. 
 
The goal for this work is to develop a fully open-source code-base for monostatic pulsed backscatter meteor radars to measure pre-$t_{0}$ speeds with realistic uncertainties.  This is partially motivated by our desire to design a fully-automated, independent meteor speed algorithm to allow for quality control of the TOF velocity solutions generated by CMOR. Although CMOR TOF speeds are generally considered to be quite accurate, having been compared to literature meteor shower speeds \cite{Brown_etal_2008} and simultaneous optical measurements \cite{Weryk_Brown_2012}, their uncertainties are affected by echo signal-to-noise ratio (SNR) and the sampling rate of the radar. The lowest relative uncertainty range for TOF speeds with CMOR occurs between speeds of about 20 km/s and 40 km/s. Below 20 km/s, the ionization efficiency drops off rapidly, causing detected echoes of fixed mass to have lower SNR and fewer remote station detections. Above 40 km/s, the comparatively low pulse-repetition frequency (PRF) of CMOR (532 Hz) results in larger errors due to the smaller time offsets between stations at higher speeds. Having an independent speed determination for these lower speed echoes, in particular, allows us to filter out echoes which have larger speed uncertainty. We test the developed algorithm by applying it to meteor backscatter echoes measured by the multi-station Canadian Meteor Orbit Radar (CMOR) and to the Middle Atmosphere Alomar Radar System (MAARSY). Both of these systems have independent speed algorithms which allow \change[R2]{us to compare}{comparison} to the proposed speed algorithm.

\section{Transverse Meteor Echo Scattering Theory}

When a meteoroid approaches the Earth, it encounters an atmosphere that exponentially increases in mass-density as altitude decreases. As the meteoroid heats up and begins to ablate, it becomes visible and can be detected optically. Meanwhile, collisions of the ablated material with air molecules produce ions and free electrons in the meteor trail which scatter radio energy. This ionized plasma, under the right geometrical conditions, can be detected if radar pulses are transmitted and the energy reflected from the trail electrons is recorded at receivers on the ground.  These detections are then used to determine the pre-Earth encounter heliocentric orbit of the meteoroid.

For the simplest case of a co-located transmitter and receiver, there is a point on the meteor trail where reflected energy will scatter in phase to the receiver along its original path. This \change[R2]{point is known as }{is }the \change[R2]{$t_0$ or specular point}{specular point corresponding to $t_0$}. Geometrically, the meteor's path must be orthogonal to the transmitted wave at this point (Figure \ref{fig:TOFgeom}). \change[R2]{Since the angle of incidence of the incoming  wave is equal to the angle of reflection from the meteor trail, there also exist other geometries for which forward scattered energy can be received by remote receiving stations - each having their own $t_n$ point}{There also exist other geometries for which forward scattered energy can be received by remote receiving stations - each having their own $t_n$ point - where the angle of incidence of the incoming wave is equal to the angle of reflection from the meteor trail}. It is this more complex arrangement that forms the basis of TOF solutions \cite{Jones_etal_2005}. Each individual electron (in the idealized underdense limit) at the specular point scatters the incoming radio waves independently, and each electron \add[R2]{at the specular point }re-radiates back to the \change[R2]{receiver}{antenna} a power:

\begin{linenomath*}
\begin{equation}
   P_{R_e} =\frac{P_T G_T G_r \lambda^2 \sigma_e}{64\pi^3 R_o^4}
   \label{singleelectron}
\end{equation}
\end{linenomath*}
%%%%%
where \remove[R2]{$P_{R_e}$ is the power received at the antenna from a single electron at the specular point, }P$_T$ is the peak transmit power, G$_T$ and G$_r$ are the antenna gain to the specular point for the transmit antenna and receiver respectively \cite{Ceplecha_1998}. Here the radar wavelength is given as $\lambda$, the classical electron back-scattering cross section $\sigma_e$ is 10$^{-28}\mathrm{~m}^2$ and the specular range is $R_o$. All electrons near the specular point scatter in phase provided the total difference in range between the segment of the trail and the receiver is less than $\frac{\lambda}{2}$. It is straightforward to show \cite<e.g.,>[]{Lovell1948, mckinley1961, Ceplecha_1998} that an extension of Eq. (\ref{singleelectron}) to the total power received is an integral summation of the long line of electrons in this idealized underdense trail which produces:

\begin{linenomath*}
\begin{equation}
   P_{R} =\frac{P_T G_T G_r \lambda^3 q^2 {r_e}^2}{64\pi^2 R_o^3}\times\left[\mathcal{F}_\mathrm{cos}^2+\mathcal{F}_\mathrm{sin}^2\right]
\label{eq:totalpower}
\end{equation}
\end{linenomath*}  \note[Gunter]{the equation seems to different by a factor to the first one, it should be 16 pi*pi, instead of 64 pi*pi, please check}

where the leading constant term contains ${r_e}$, the classical electron radius, $\sqrt{\sigma_e/(4\pi)}$ and $q$ is the number of electrons per unit path length of the trail. The term in square brackets represents the time-varying component of the received signal in the form of Fresnel integrals. This is equivalent to the problem of knife-edge diffraction in physical optics.
Since the ionized meteor trail is typically several to tens of kilometres long, \change[R1]{it}{a meteor} crosses multiple Fresnel zones. As a result, the amplitude and phase components of the received signal will have characteristic oscillations as the meteor crosses the alternating zones of constructive and destructive interference described by near-field Fresnel diffraction \cite{mckinley1961}. The length of any Fresnel zone, $F_n$, can be approximated by,
%%%%%
\begin{linenomath*}
\begin{equation}
    F_n = \sqrt{\frac{n\lambda R_1 R_2}{R_1+R_2}}
\end{equation}
\end{linenomath*}
%%%%%
where $n$ is the zone number, $\lambda$ is the wavelength of the energy being scattered, $R_1$ is the transmitter-$t_n$ distance, and $R_2$ is the $t_n$-receiver distance. Assuming a co-located transmitter and receiver ($R_1 = R_2$) with a radar wavelength of 10 m for a meteor with a range of 100 km, the first five Fresnel zones will have \change[R2]{radii}{distances} from the $t_0$ point of: $F_1=707$ m, $F_2=1000$ m, $F_3=1225$ m, $F_4=1414$ m, and $F_5=1581$ m. To qualitatively understand the amplitude and phase behaviour of the energy scattered from these zones, the Fresnel integrals in Eq. \eqref{eq:totalpower} are given as Eqs. (\ref{eq:fsin} and \ref{eq:fcos}).
%%%%%
\begin{linenomath*}
\begin{equation}
    \mathcal{F}_\mathrm{sin} = \int_{-\infty}^{x} \sin\bigg(\frac{\pi}{2}x^2\bigg) dx
\label{eq:fsin}
\end{equation}
\end{linenomath*}
%%%%%
\begin{linenomath*}
\begin{equation}
     \mathcal{F}_\mathrm{cos} = \int_{-\infty}^{x} \cos\bigg(\frac{\pi}{2}x^2\bigg) dx
\label{eq:fcos}
\end{equation}
\end{linenomath*}
%%%%%
where $x$ is the Fresnel parameter or length along the spiral formed by the parametric plot of the Fresnel integrals.

\begin{figure}
    \centering
    \includegraphics[width=\linewidth]{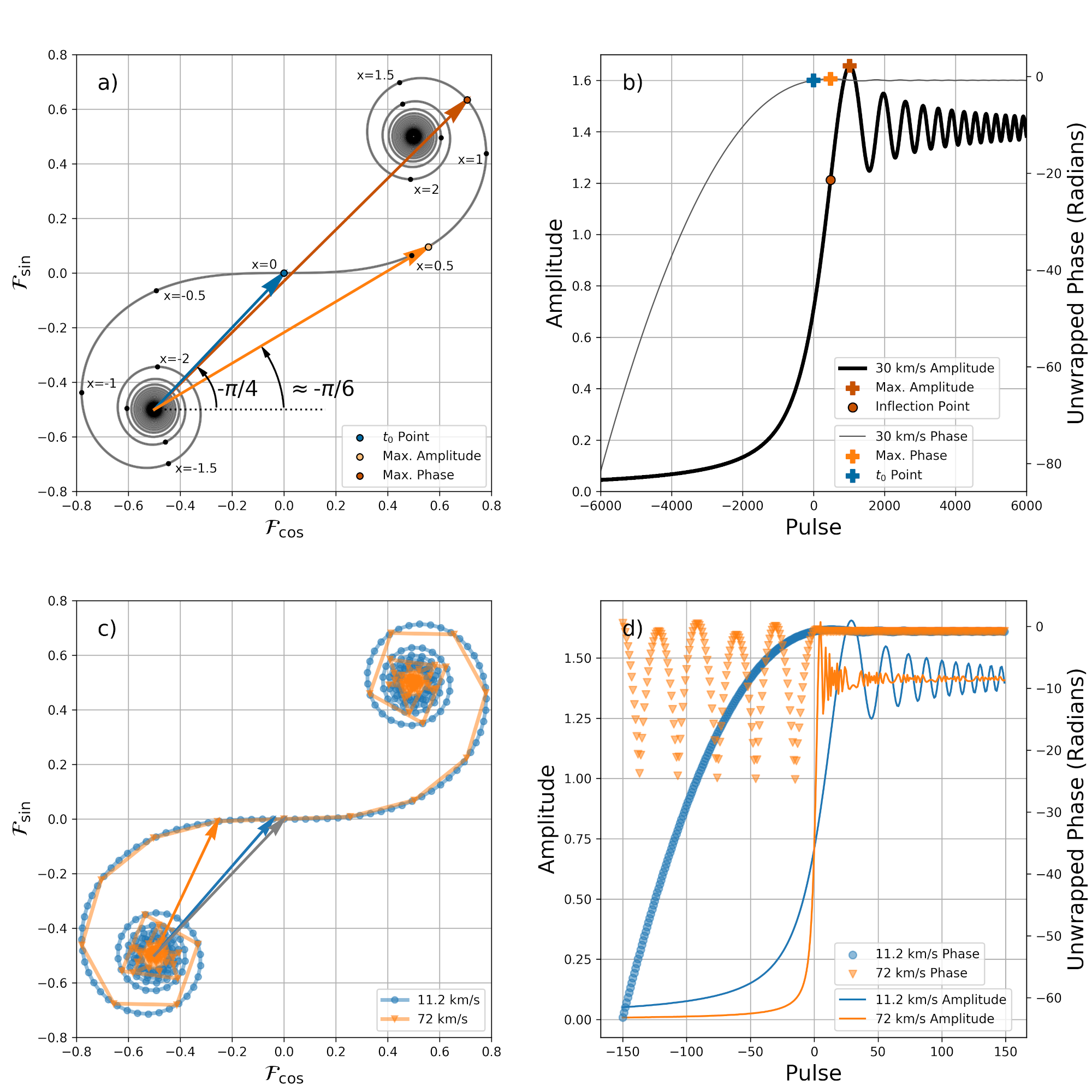}
    \caption{a) Model backscatter meteor echo Fresnel integral plot. b) Echo amplitude and phase calculated from a). c) Cornu spirals for echoes with speeds of 11.2 and 72 km/s at a sample rate of 532 Hz. d) Amplitude and phase for the echoes described by c). In all plots, the \change[petr]{t0}{$t_0$} point occurs where the curves cross x=0.}
    \label{fig:fresnelsimulation}
\end{figure}

A plot of $\mathcal{F}_\mathrm{sin}$ against $\mathcal{F}_\mathrm{cos}$ produces the Cornu Spiral (Panels a) and c) in Figure \ref{fig:fresnelsimulation}) which can be used to visualize the amplitude and phase behaviour of an echo as a function of time. In \change[mazur]{Plot}{plot} a)\add[mazur]{,} the \add[R1]{amplitude of the meteor }echo can be thought of as beginning at the (-0.5,-0.5) point at a distance of $-\infty$ and spiralling outwards along the curve until it reaches the $t_0$ point at (0,0). As it does this, the length of the vector from the $-\infty$ point to its location on the spiral increases. This represents the growth in the amplitude of the echo. The phase of the echo at any point is the angle from the horizontal line through the $-\infty$ point to its location on the spiral. \add[mazur]{From this, it follows} that the amplitude of the echo at the $t_0$ point will be $\sqrt{2*0.5^2}$ while the phase will be $-\pi/4$. Beyond the $t_0$ point, the phase will increase to a maximum of approximately -0.513 radians (-29.4$^\circ$) where it is tangent to the spiral at $x=0.572$ before it decreases, passes through the maximum amplitude of 1.657 at $x=1.217$, and then oscillates around $-\pi/4$ as it moves towards $\infty$. \add[mazur]{It can also be observed that the rate of change in amplitude begins to decrease at the point of maximum phase. On plot b) this is the amplitude inflection point that is used for correlating CMOR echoes in the TOF method described in Appendix A. }Note that this treatment ignores the effects of plasma resonance \cite{Kaiser1952} which may change the phase by up to 180$^\circ$, depending on the orientation of the trail relative to the plane of the transmitted electric field. 

In general terms, the time varying component of the amplitude $\mathcal{A}$ of the echo is,
%%%%%
\begin{equation}
    %\sqrt{\pow{\Delta fcos}{2} + \pow{\Delta fsin}{2}}
    \mathcal{A}=\sqrt{(\Delta \mathcal{F}_\mathrm{cos})^2 + (\Delta \mathcal{F}_\mathrm{sin})^2}
\label{eq:echoamp}
\end{equation}
%%%%%
while the phase is given by,
\begin{equation}
    \phi = \arctan\bigg(\frac{\Delta \mathcal{F}_\mathrm{sin}}{\Delta \mathcal{F}_\mathrm{cos}}\bigg)
\label{eq:echophs}
\end{equation}
%%%%%
where $\Delta \mathcal{F}_\mathrm{cos}$ and $\Delta \mathcal{F}_\mathrm{sin}$ are measured from (-0.5,-0.5) in the Cornu spiral plot.

\begin{figure}
    \centering
    \includegraphics[width=\linewidth]{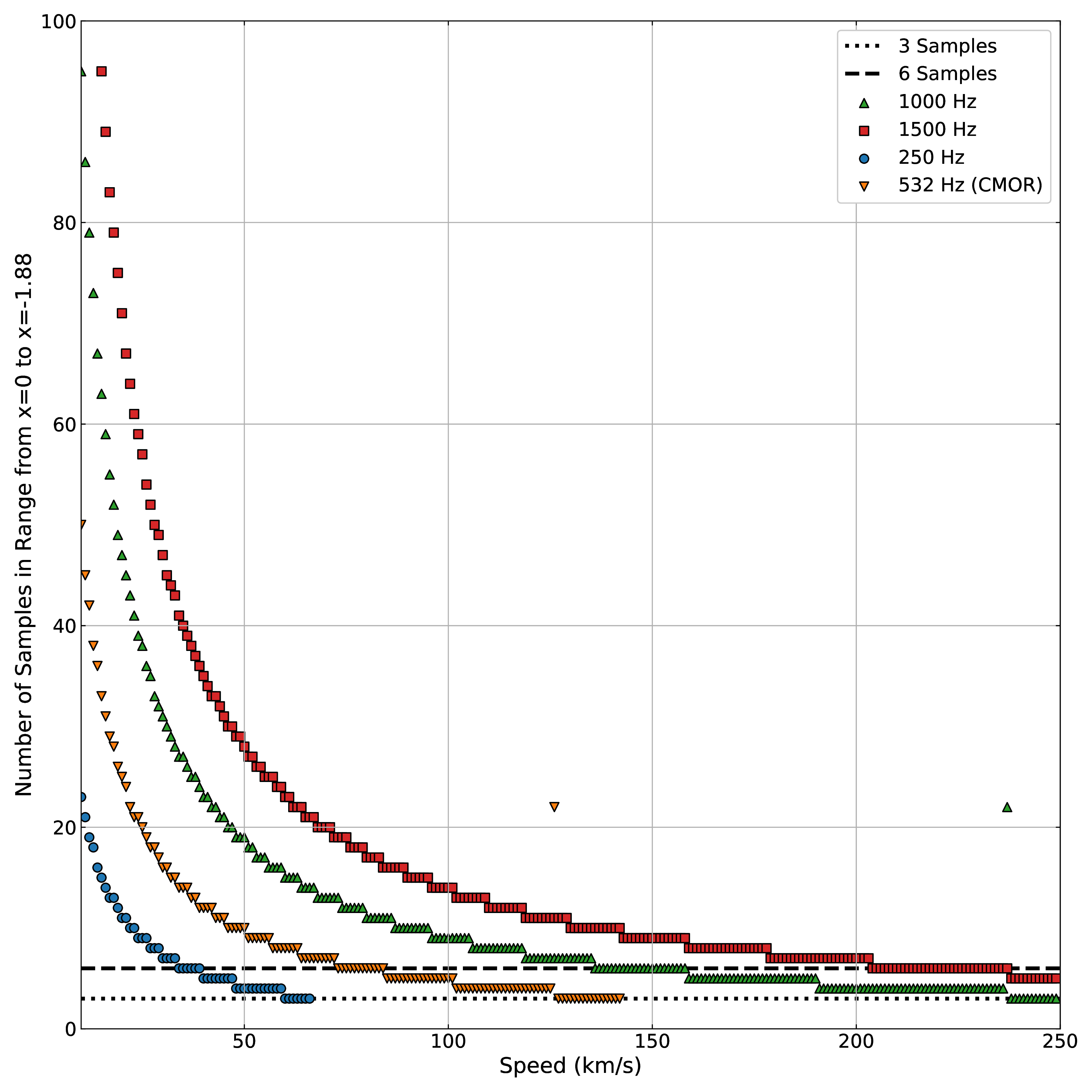}
    \caption{The number of pulses (samples) in the first 2$\pi$ phase increment after the $t_0$ ($x=0 \rightarrow x=-1.88$) \change[R1]{vs.}{versus} meteoroid speed. Here the range is 100 km and the radar wavelength is 10 m in all cases. \add[R2]{Although a minimum of three samples are typically required, aliasing can occur when the number of samples changes from four to three. Three (absolute) and six (practical) sample aliasing limits are shown by the dotted and dashed lines respectively.}}
    \label{fig:PRFsensitivity}
\end{figure}

Plots a) and b) of Figure \ref{fig:fresnelsimulation} describe an idealized model echo sampled at a rate of 50 kHz. For real radars, the pulse repetition frequency (PRF) is much lower and this adds significant speed-dependent uncertainties. Plot c) of Figure \ref{fig:fresnelsimulation} shows the difference between echoes having speeds of 11.2 km/s and 72 km/s sampled at the CMOR PRF of 532 Hz. The grey arrow points to the $t_0$ point while the other two arrows point to the first sample before $t_0$ for each speed. This shows that, for low speed echoes, the uncertainty in phase and amplitude will be much less than for high speed echoes. Plot d) in Figure \ref{fig:fresnelsimulation} shows that the 11.2 km/s echo has smooth, nearly continuous phase and amplitude profiles while the phase for the 72 km/s echo has not been properly unwrapped by a simple, 2$\pi$ point-to-point shift method. As the speed continues to increase, the number of samples per 2$\pi$ change in phase decreases and the result is aliased. Figure \ref{fig:PRFsensitivity} shows the number of samples in the first 2$\pi$ rotation of phase before $t_0$ ($x=0 \rightarrow x=-1.88$) for four PRFs from 250 Hz to 1500 Hz. \change[mazur]{In practice}{Although 3 samples is the absolute limit}, we find that\add[mazur]{, in practice,} when the number of samples is fewer than about 6, the risk of aliasing during phase unwrapping increases significantly. For PRFs of 250 Hz, 532 Hz, 1000 Hz, and 1500 Hz, this occurs at 40 km/s, 85 km/s, 159 km/s, and 238 km/s respectively. \remove[mazur]{In real echoes with noise, echoes with $\sim$ 10 pulses is roughly where we see issues with aliasing begin.}

Following the convention of \citeA{Baggaley_etal_1997}, the distance of the meteoroid from $t_0$ is,
%%%%%
\begin{equation}
    s = x \frac{\sqrt{R\lambda}}{2}
\label{eq:distancealongtrail}
\end{equation}
%%%%%
where $s$ is the linear distance along the trail from the $t_0$ point and $R$ is the range to the echo. It follows, then, that the speed of the meteoroid is,
%%%%%
\begin{equation}
    v = \frac{\sqrt{R\lambda}}{2} \frac{dx}{dt}
\label{eq:speed}
\end{equation}
%%%%%
As such, the speed of a meteoroid can be found graphically by measuring the slope of the pre-$t_0$ echo signal in an \change[R2]{s}{$s$} \change[R1]{vs.}{versus} \change[R2]{t}{$t$} plot. It is this equation that forms the basis of the pre-$t_0$ method that follows.

\section{Method}
\label{sec:meth}

Since 2002, CMOR has recorded more than 15 million multi-station orbits. In addition, there are upwards of 50 million single-station echoes that have not been fully solved for speed. The algorithm described here, which we term the pre-$t_{0}$ speeds by sliding-slopes technique (PSSST) has been developed for CMOR as part of a project to better understand the low-speed meteoroid population observed by the system. Because of CMOR's multi-station configuration, analysis of data from CMOR has relied on geometric time-of-flight (TOF) speeds (see Appendix A for details). With a pulse repetition frequency (PRF) of 532 Hz (Table \ref{tab:CMORspecs}) large uncertainties in TOF solutions can occur for some geometries where specular points are not well-separated along the trail, particularly at high speeds. At low speed, the bulk of the TOF speed uncertainty is from weak echo amplitudes and correspondingly poor inflection point selection (see also Appendix A for the inflection point algorithm). Since reliable phase information is still present at low amplitudes \cite{Cervera_etal_1997}, the pre-$t_{0}$ method provides good speeds for echoes that tend to produce uncertain TOF solutions. In an attempt to quantify uncertainty in the speed estimate for low and high speed meteors, a robust estimation method has been developed extending the pre-$t_{0}$ method first developed by \citeA{Cervera_etal_1997}.

\begin{table}
    \caption{Equipment and experiment waveform specifications of CMOR and MAARSY}
    \centering
    \begin{tabular}{l c c}
    \hline
    & CMOR & MAARSY\\
    \hline
    Location (Lat., Long.) & 43.264$\degree$N, 80.772$\degree$W & 69.30$\degree$N, 16.04$\degree$E \\
    Frequency & 29.85 MHz & 53.5 MHz \\
    Pulse repetition frequency & 532 & 1000 \\
    Range sampling interval & 15-252 km & 69.75-139.95 km\\
    Peak transmitter power & 12 kW & 800 kW\\
    Range Resolution & 3 km & 0.45 km\\
    Beam width (3dB) & 30$\degree$ & 3.6$\degree$\\
    Beam pointing & Vertical & 30$\degree$ off zenith \\
    \hline
    \end{tabular}
    \label{tab:CMORspecs}
\end{table}

\begin{figure}
    \centering
    \includegraphics[width=\linewidth]{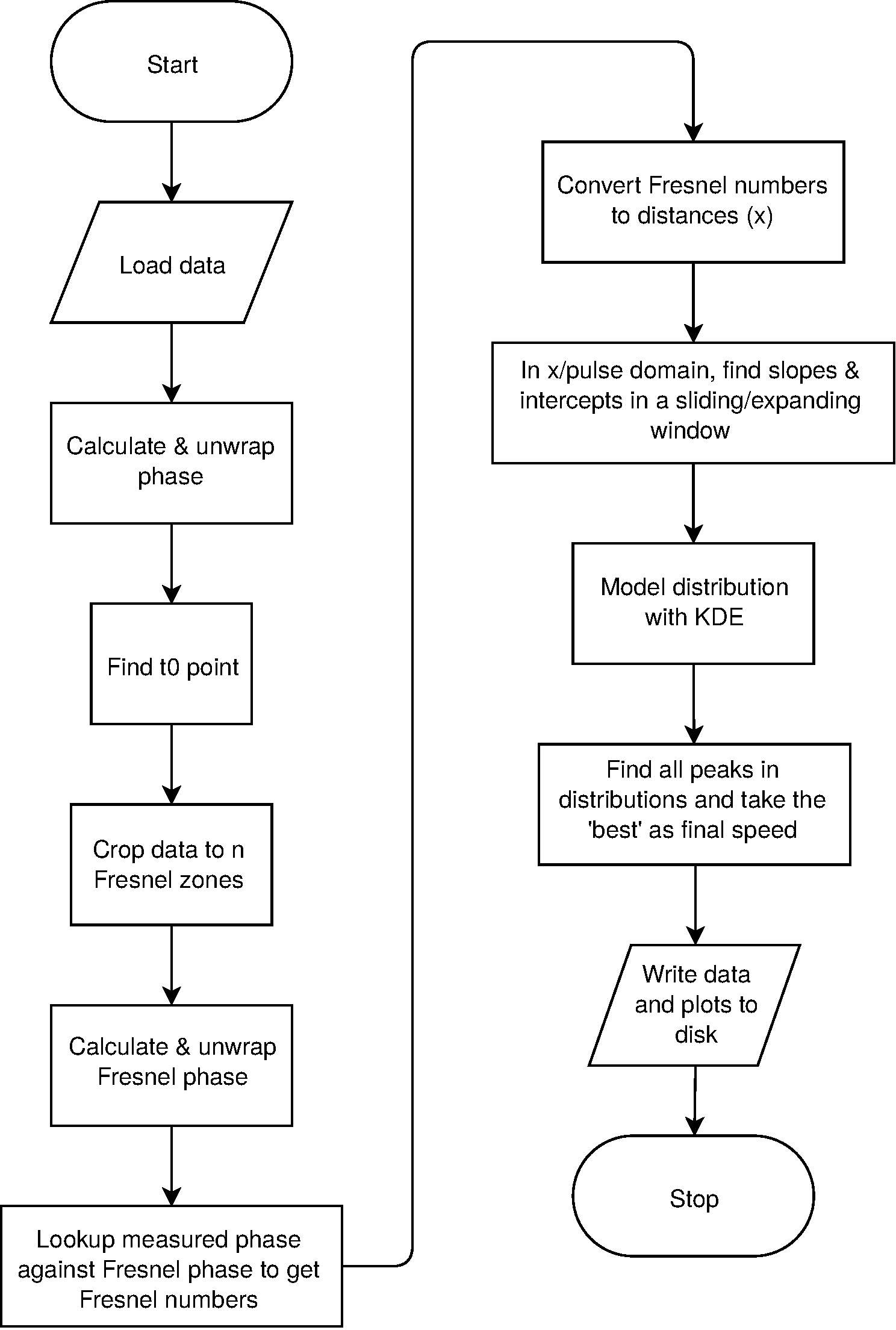}
    \caption{Workflow for computing pre-$t_{0}$ speeds by the sliding-slope technique.}
    \label{fig:flowchart}
\end{figure}

The basic algorithm (Figure \ref{fig:flowchart}) consists of the following steps: pre-processing, Fresnel distance lookup, speed determination, and plotting. Pre-processing consists of loading the data, calculating phase and amplitude, smoothing the amplitude and unwrapped phase, detrending for winds, and finding the $t_{0}$ point. Phase unwrapping is done by a sequential point-scanning method (SPSM). The phase of the received signal as a function of time is scanned and whenever a 2$\pi$ discontinuity is encountered, the same 2$\pi$ shift is applied. Other, more complicated methods were tried \cite{estrada2011noise, xu2013robust}, but SPSM was found to be the most robust.

The effect of upper middle-atmospheric winds on the observed phase can change the measured speed by the order of 1 km/s or more, and must be removed prior to finding the $t_{0}$ point. The method used here \citeA<following>[]{Cervera_1996} assumes that the slope of the phase curve post maximum amplitude, is proportional to the radial velocity of the wind acting on the meteor trail. For CMOR, the slope is measured within a 50 pulse window after maximum amplitude using a robust RANdom SAmple Consensus (RANSAC) regressor \cite{Fischler_Bolles_1981} to minimize the effects of outliers. This Doppler shift is given by \note[Gunter]{ the definition of the Doppler velocity comes with a minus to account for the phase delay dphi/dt=-4Pi/lambda vr},
%%%%%
\begin{linenomath*}
\begin{equation}
    \Delta f = \frac{d \phi}{dt}.
\end{equation}
\end{linenomath*}
%%%%%
From this, the wind speed, $v_\mathrm{rad}$, in m/s is simply,
%%%%%
\begin{linenomath*}
\begin{equation}
    v_\mathrm{rad} = \add[R2]{-}\Delta f \frac{\lambda}{2},
    \label{eq:windspeed}
\end{equation}
\end{linenomath*}
%%%%%
where $\lambda$ is the wavelength of the radar in meters, and $\Delta f$ is measured in \change[R1]{cycles/s}{Hz}. Following the notation of \citeA{Cervera_1996}, detrending of the complex wind-contaminated signal, $E^\prime$, is accomplished by applying the following transformation.
%%%%
\begin{linenomath*}
\begin{equation}
    E = E^\prime e^{\add[R2]{-}i {\Delta f} t},
\end{equation}
\end{linenomath*}
%%%%
where $t$ is time measured from some arbitrary zero-point (we choose the maximum amplitude point). 

As a check of the algorithm's \add[R1]{line-of-sight, }neutral wind estimation method, we compare the results for a day of CMOR data using Eq. \eqref{eq:windspeed} with the wind speeds found using the technique described by \citeA{Hocking_etal_2001} which is generated automatically by the SKiYMET cross/auto-correlation algorithms which run in parallel with the CMOR meteor analysis routines. The SKiYMET neutral wind speeds generally have uncertainties of less than 1 m/s and are widely used in the aeronomy community \cite<e.g.,>[]{Stober_etal_2014}. 

\begin{figure}
    \centering
    \includegraphics[width=\linewidth]{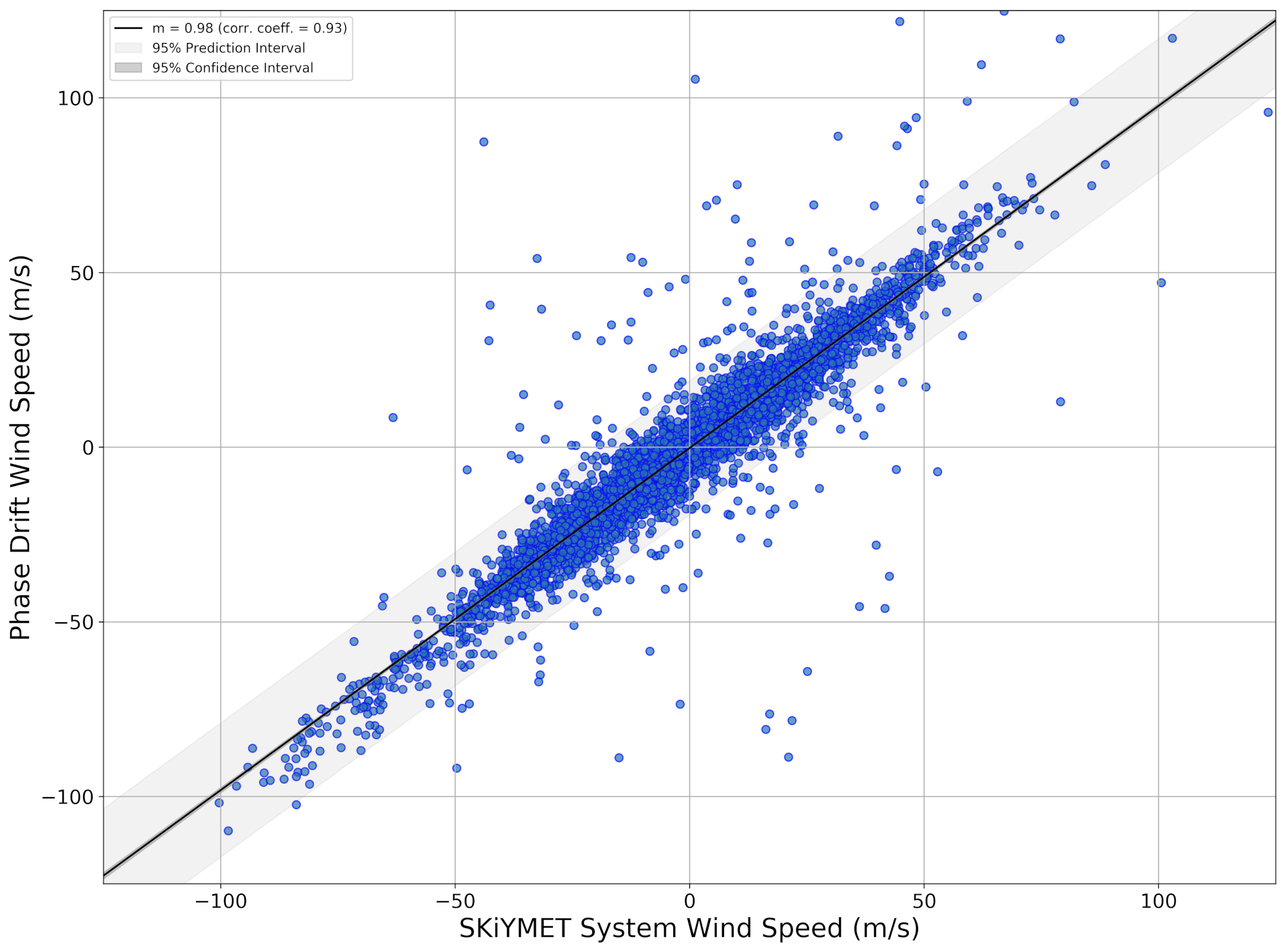}
    \caption{\change[R1]{W}{Radial, line-of-sight w}ind speeds calculated by the phase-drift method described in this paper are compared to wind speeds for the same 4775 echos automatically calculated by the SKiYMET system following the technique described in \citeA{Hocking_etal_2001}. These are from CMOR echoes collected on October 8, 2012.}
    \label{fig:windspeedcomp}
\end{figure}

However, we only have CMOR SKiYMET wind speeds for a fraction of the total echoes observed, due to the strict acceptance criteria by the native SKiYMET software used for wind measurements. Comparison of \add[R1]{radial }wind speeds from our technique  against the SKiYMET-determined wind speed (Fig. \ref{fig:windspeedcomp}) shows that there is relatively good agreement between the two with little scatter from a 1:1 relation \add[mazur]{(slope = 0.98, correlation coefficient = 0.93)}. This suggests that our correction is adequately removing the effects of wind drift from pre-$t_{0}$ speeds.

The magnitude of the effect that winds have on PSSST speeds is made apparent when echoes with and without wind corrections at different wind speeds are examined (Table \ref{tab:windcorrnowindcorr}). Note that the TOF speed uncertainties are computed using a Monte Carlo error estimation technique described by \citeA{Weryk_Brown_2012}. For low wind speeds, there is only a small effect on the speed estimation but, for high speeds, the winds can cause PSSST speeds to be inaccurate on the order of 1 km/s.

\begin{table}
    \caption{Three example CMOR echoes showing the effects of \add[R1]{radial }wind speed corrections on PSSST speeds\add[mazur]{$v_{Pw}$ and $v_{P}$ are the PSSST speeds with and without wind corrections respectively. Asymmetric error bounds result from the KDE error estimation method.}\note[mazur]{Changed column headings.}}
    %PP: I added \add[mazur] so it is reflected as a change you made
    \centering
    \begin{tabular}{l c c c c}
    \hline
    \multicolumn{1}{c}{$v_{TOF}$} & $v_{wind}$ & $v_{P}$$^{+error}_{-error}$ ($\Delta$$^{a}$) & $v_{Pw}$$^{+error}_{-error}$ ($\Delta$$^{a}$)\\
    \multicolumn{1}{c}{[km s$^{-1}$]} & [m s$^{-1}$] & [km s$^{-1}$] & [km s$^{-1}$]\\
    \hline
    11.28 $^{+}_{-}$ N.A.$^{b}$ & 0.46 & 11.00 $^{+0.22}_{-0.23}$ (-0.28) & 11.02 $^{+0.22}_{-0.23}$ (-0.26)\\
    17.48 $\pm$1.12 & -25.29 & 16.95 $^{+0.36}_{-0.50}$ (-0.53) & 17.96 $^{+0.71}_{-0.67}$ (+0.48)\\
    30.44 $\pm$0.28 & 46.85 & 31.65 $^{+0.89}_{-0.92}$ (+1.21) & 30.23 $^{+0.48}_{-0.61}$ (-0.21)\\
    \hline
    \multicolumn{4}{l}{$^{a}$$\Delta$ is the difference between the PSSST speed and the measured TOF speed.}\\
    \multicolumn{4}{l}{$^{b}$TOF error was not calculable for this echo.}
    \end{tabular}
    \label{tab:windcorrnowindcorr}
\end{table}

Knowledge of the $t_{0}$ point is required for transforming the phase data to distance along the trail. Although our algorithm typically estimates $t_{0}$ to within about $\pm$2 pulses, our $t_{0}$-dependent weighting and minimum phase normalization scheme can result in a poor $t_{0}$ pick and associated errors of many km/s in speed. To find the $t_{0}$ point, wind effects are first removed and the maximum echo amplitude is taken as a first guess at $t_{0}$, knowing that this should be well after the real $t_{0}$ point for normal echoes. The algorithm then moves back in time from this pulse until the point of maximum unwrapped phase. The entire phase vs. time is then shifted \add[R2]{in phase }such that the maximum phase coincides with the model expected -29.4$\degree$. The algorithm then moves back further, with the $t_{0}$ point declared to be found when the phase equals -$\pi$/4. The temporal analysis window is then set to a number of Fresnel zone widths before and after the $t_{0}$ point, which we found for CMOR to be best fixed to six. Finally, the unwrapped phase curve is compared to the expected Fresnel phase using Fresnel integrals, \remove[R1]{ultimately} generating a look-up table converting unwrapped phase to distance along the trail. 

For each pulse, the  distance (from the $t_{0}$ point) corresponding to a phase measurement at a given time is determined using Eq. (\ref{eq:distancealongtrail}). From the $s$ values, point-to-point slopes and intercepts are then computed within a sliding, dynamically-sized window in time along a six Fresnel zone wide section of the signal prior to the $t_{0}$ point. A weighted histogram is then created to describe the distribution of measured speeds and a Kernel Density Estimator (KDE) \cite{rosenblatt1956, parzen1962} applied to estimate the best speed and uncertainty. Although choice of KDE basis functions and bandwidth can greatly affect the shape of the KDE \cite{Vida2017}, we've assumed Gaussian distributions and let the bandwidth be chosen by Scott's rule-of-thumb \cite{Scott2012MultivariateDE}.

The advantage of this approach is that if the $t_{0}$ pick is incorrect by of order a couple of pulses the speed will still be close to the true value as the approach generates all possible speeds from line segments of length Y to Y-N along the distance vs. time curve and not a single estimate from one slope reliant on a correct choice of t$_{0}$. 

\begin{figure}
    \centering
    \includegraphics[width=\linewidth]{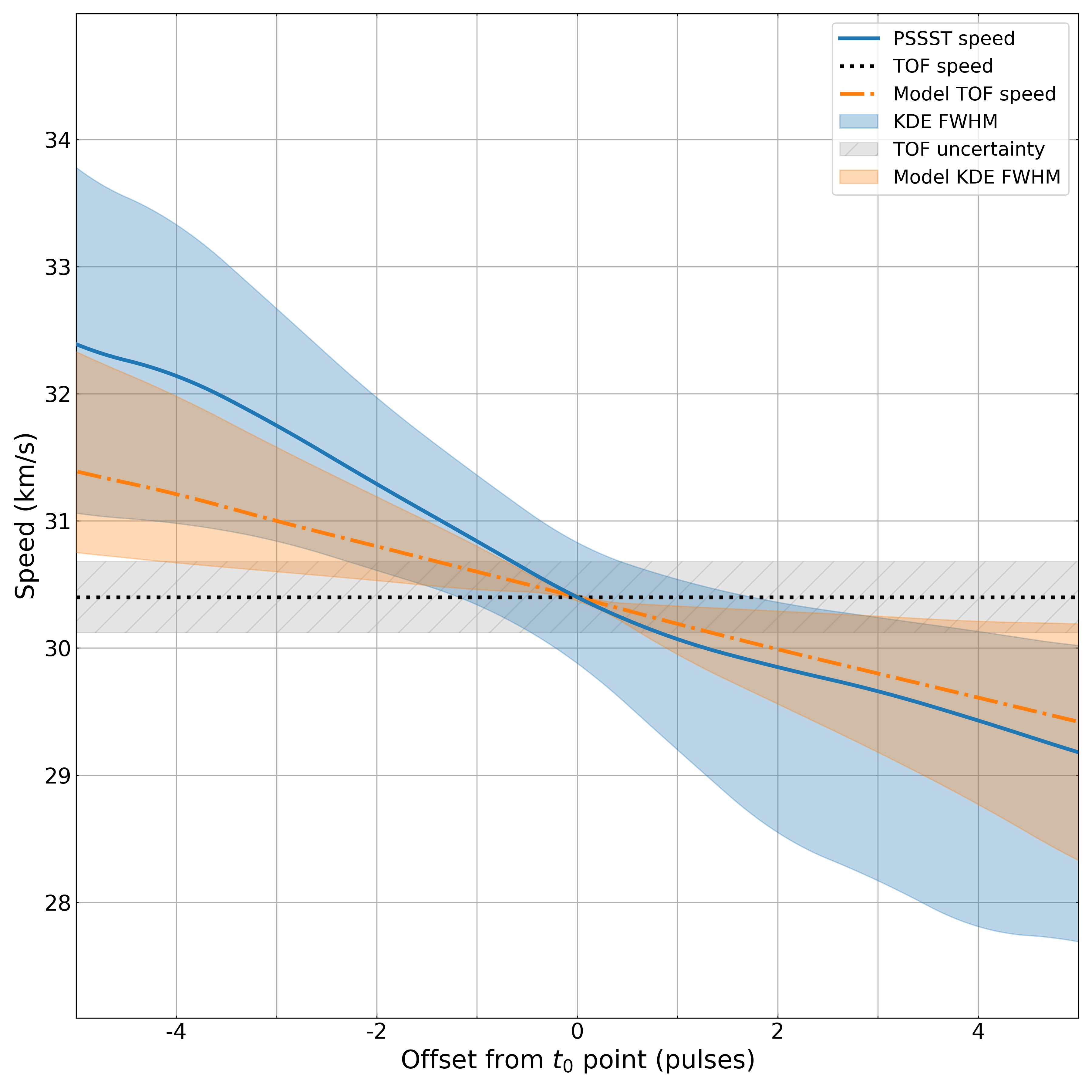}
    \caption{The PSSST echo speed sensitivity to offset from the automated $t_{0}$ pick location for a real echo with a TOF speed of 30.44$\pm$0.28 km/s (solid line). The true $t_{0}$ location is located at the KDE FWHM minimum and is consistent with the auto-pick. Also shown are the results for a noise-free synthetic echo with the same speed (dash-dot line).\add[R2]{ Here we define a negative offset as earlier than the auto-pick $t_0$ point and a positive offset as later.}}
    \label{fig:t0sensitivity}
\end{figure}

\begin{figure}
    \centering
    \includegraphics[width=\linewidth]{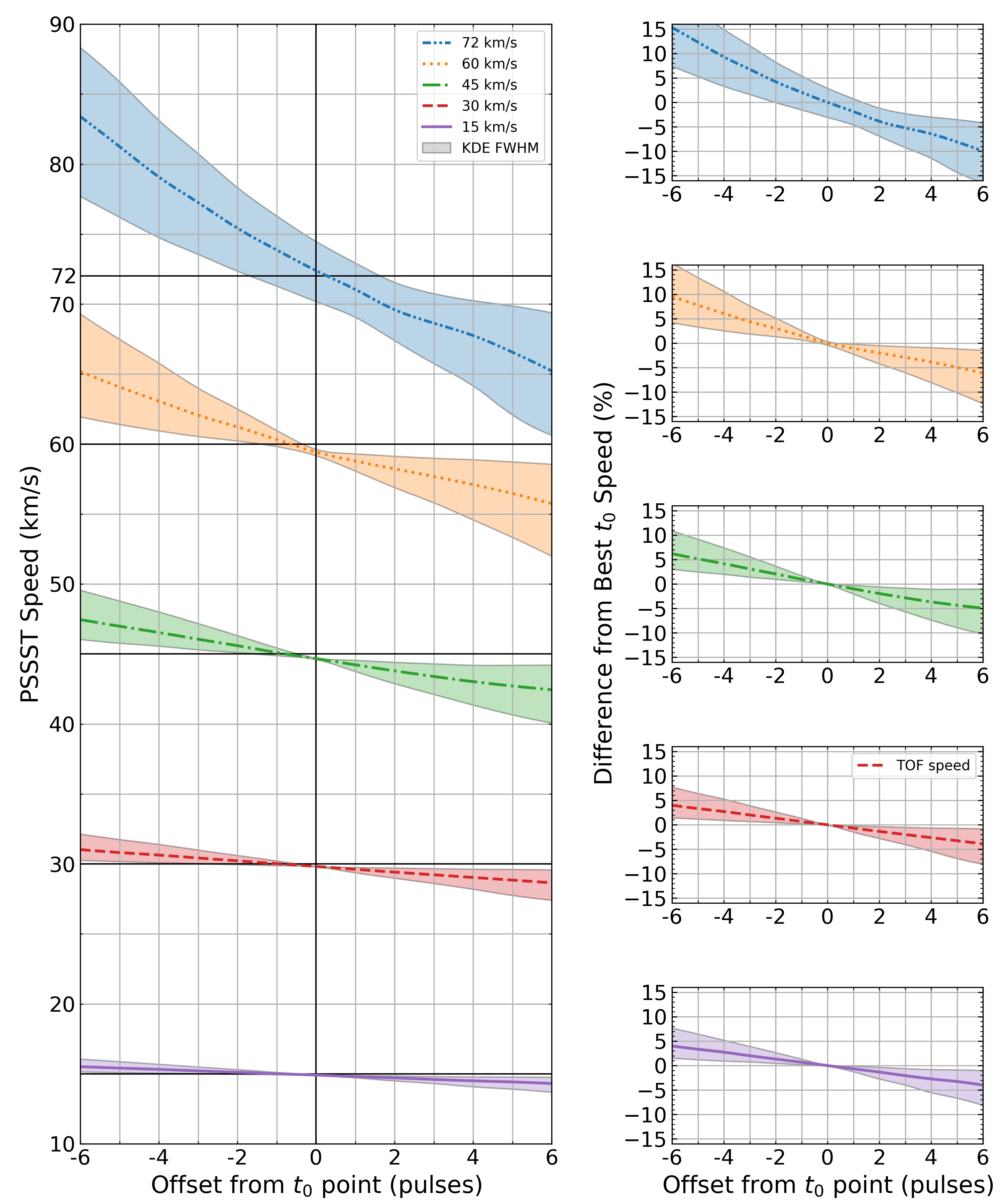}
    \caption{Sensitivity of the $t_{0}$ point pick on PSSST speeds shown for synthetic and noise-free echoes at true speeds of 15, 30, 45, 60, and 72 km/s, respectively. The left panel shows the PSSST results at different \change[petr]{t0}{$t_0$} offsets, while the right panels show the percentage differences as the \change[petr]{t0}{$t_0$} point is varied. Here we define a negative offset as earlier than the auto-pick \change[petr]{t0}{$t_0$} point and a positive offset as later.}
    \label{fig:t0modelsensitivity}
\end{figure}

Figure \ref{fig:t0sensitivity} shows an example echo and how the PSSST speed can vary with $t_{0}$ offset from the auto-picked location. The full-width, half maximum (FWHM) of the KDE is minimum at the auto-pick location and increases as the $t_{0}$ point is forced earlier (-ve offset) or later (+ve offset) in time. In this case, an offset of +2/-1 pulses would still give a PSSST speed that is within the KDE FWHM uncertainty. Figure \ref{fig:t0modelsensitivity} shows $t_{0}$ sensitivity for five noise-free synthetic echoes at speeds of 15, 30, 45, 60, and 72 km/s. In the left-hand panel, both the deviation from the true speed and the uncertainty can be seen to increase with speed. Additionally, the PSSST speed is typically slightly less than the model speed with the best pick being about 0.5 pulses earlier in time. Although the algorithm tries to pick the integral pulse closest to $t_0$, further improvements could be made by simply interpolating the phase and amplitude prior to $t_0$ picking. In terms of the percentage difference from the auto-pick speed, a\change[R1]{ mispick}{n error in the chosen $t_0$ point} as small as 3 pulses can result in \add[R1]{speed }errors \add[R1]{of }between 2$\%$ (15 km/s) and 7$\%$ (72 km/s) \add[R2]{(right-hand panels)}. A 6 pulse \change[R1]{mispick}{$t_0$ picking error} may result in \add[R1]{speed }errors \add[R1]{of }between 4$\%$ and 15$\%$\remove[R2]{ (right-hand panels)}. For speeds less than about 50km/s, the $t_0$ point can be wrong by as much as $\pm$6 pulses and still give results that are within $\pm$5$\%$ of the \change[R2]{'}{"}true\change[R2]{'}{"} speed. At the highest speeds, however, the $t_0$ pick becomes more critical, with $\pm$2 pulse picking tolerance required for $\pm$5$\%$ accuracy. 

In addition to using the KDE width as a measure of $t_{0}$ pick accuracy, it should also be possible to use the shape of the KDE to constrain the direction of the offset. As the $t_{0}$ point is offset from the true location, the shape of the KDE skews towards the true speed. This is apparent in Figure \ref{fig:t0sensitivity} where, for the positive offsets, the PSSST speed line does not track the middle of the KDE FWHM. Although it is less apparent with this example, simulations show that the KDE will also skew towards the true speed when $t_{0}$ is offset in the negative direction. A more robust algorithm could therefore be envisioned where, after the described $t_{0}$ auto-pick is made, the $t_{0}$ point is refined by minimizing the width of the KDE (or minimizing the excess kurtosis) through an iterative process that is guided by the skewness of the KDE. \add[R2]{In terms of the slope of the line describing the PSSST solutions, we see that for positive offsets, it is similar to the model echo solution. However, for negative offsets, the solutions quickly diverge from the model solution. What this tells us is that our real echo solutions are more tolerant to $t_0$ pick points occurring after the true $t_0$ point in time.}

Speeds and intercepts are separately measured and fit\add[R1]{ted} with a weighted, Gaussian KDE. The weighting scheme was chosen through trial-and-error by examining the effects of correlation coefficients, window size, and proximity to the $t_{0}$ point. \change[R1]{In the end}{As a result}, we found that weights based on correlation to the fourth power and the inverse of the distance from the window leading edge to the $t_{0}$ point was optimal. Thus we adopt a weighting of the form:
%%%%%
\begin{linenomath*}
\begin{equation}
    \mathrm{Weights} = (r - r_\mathrm{limit})^4  \Delta t_0-1
\end{equation}
\end{linenomath*}
%%%%%
where \change[R2]{r}{$r$} is the sample correlation coefficient, $r_\mathrm{limit}$ is the minimum allowed \change[R2]{r}{$r$}, and $\Delta t_{0}$  is the distance between the leading edge of the window and the $t_{0}$ point.

Each of the peaks within the KDE distributions are found, the largest being chosen as the first-estimate of the true speed (V$_\mathrm{KDE}$) and intercept values. For some low SNR echoes showing multiple peaks in the KDE, we found that the largest peak in the histogram (V$_\mathrm{peak}$) does not always represent the best speed. To address this issue, we check all possible speed/intercept pairs within $\pm$3 km/s of each peak and choose the speed/intercept solution that plots closest to the estimated $t_{0}$ point.

\begin{figure}
    \centering
    \includegraphics[width=\linewidth]{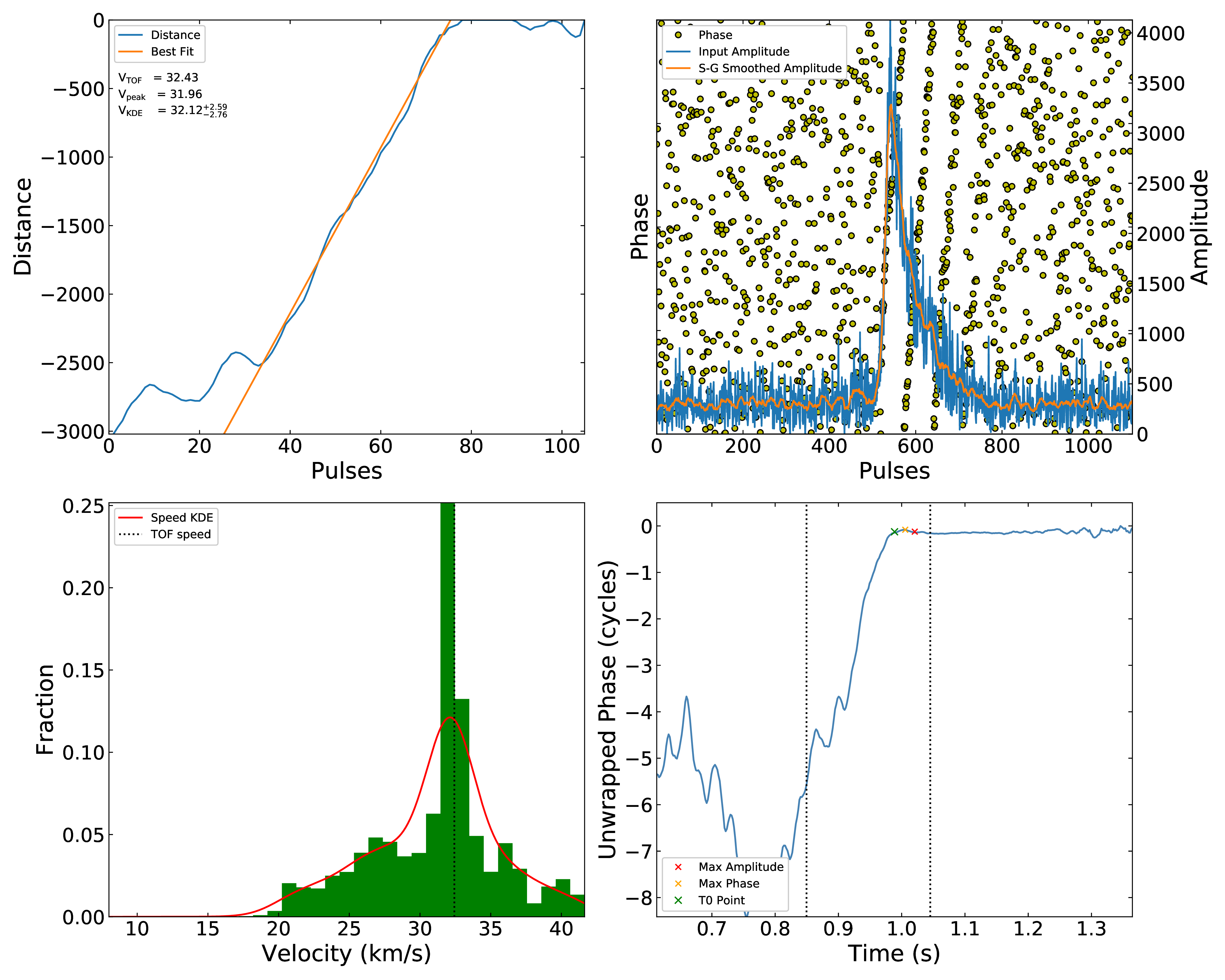}
    \caption{Example of \add[R2]{the }output from \add[R2]{the }PSSST python implementation.The upper left plot shows the Fresnel-integral distance computed from the unwrapped phase (lower right plot) and known echo range. The original echo is shown in the upper right per pulse (532 per second) with the amplitude in digital units (in blue - orange line is smoothed) and individual phase measurements per pulse as yellow circles. The lower left hand plot shows the final PSSST velocity solutions as a histogram - each velocity estimate being computed from a single sliding window of varying length combination. The red line shows the KDE applied to the histogram. For this echo, the PSSST speed is $32.12^{+2.59}_{-2.76}$ km s$^{-1}$, while the time of flight speed measured from 6 stations is 32.43 $\pm$ 64.9 km s$^{-1}$. }
    \label{fig:pssstexample}
\end{figure}

Written in Python, \texttt{Pyt0} is our implementation of PSSST as described here. It consists of about 1000 lines of code and has been used to extensively test the method prior to its incorporation into the CMOR C++ codebase.  Output consists of a number of diagnostic plots (Figure \ref{fig:pssstexample}) and text files for further interpretation of the data. The code can be directly downloaded from (https://github.com/wmpg/PSSST-Pret0).

We first verify that the PSSST algorithm performs as expected by applying it to model data. We then validate the algorithm using real data where we compare our PSSST results to Fresnel transform speeds for MAARSY transverse scattering data and to time-of-flight (TOF) speeds for CMOR transverse scattering echoes. The Fresnel transform speeds were generated using the approach of \citeA{Holdsworth_etal_2007} while the TOF solutions for CMOR are found as described by \citeA{Jones_etal_2005, Brown_etal_2008} and in Appendix A. 

\section{Algorithm Verficiation and Validation}
To verify  the algorithm, 3000 model echoes were examined in detail using PSSST and \texttt{Pyt0}. Further validation involved using  a total of nearly 12 million real echoes from two different radars. The first validation data\add[R1]{ }set (1695 echoes) was from the 53.5MHz active phased-array radar  MAARSY on And\o ya, Norway, while additional data\add[R1]{ }sets come from CMOR (12 million echoes) in southern Ontario, Canada. The meteor echoes from both systems  were processed with the PSSST algorithm and the results were compared against Fresnel transform (MAARSY) and TOF (CMOR) speeds. The pre-$t_0$ speed uncertainty is taken as the width of the KDE peak at half-maximum. We find that pre-\change[R1]{t0}{$t_0$} speeds can be computed for 98$\%$ of all CMOR echoes for which TOF speeds are measured. 

\subsection{Modeled Data}

\begin{figure}
    \centering
    \includegraphics[width=\linewidth]{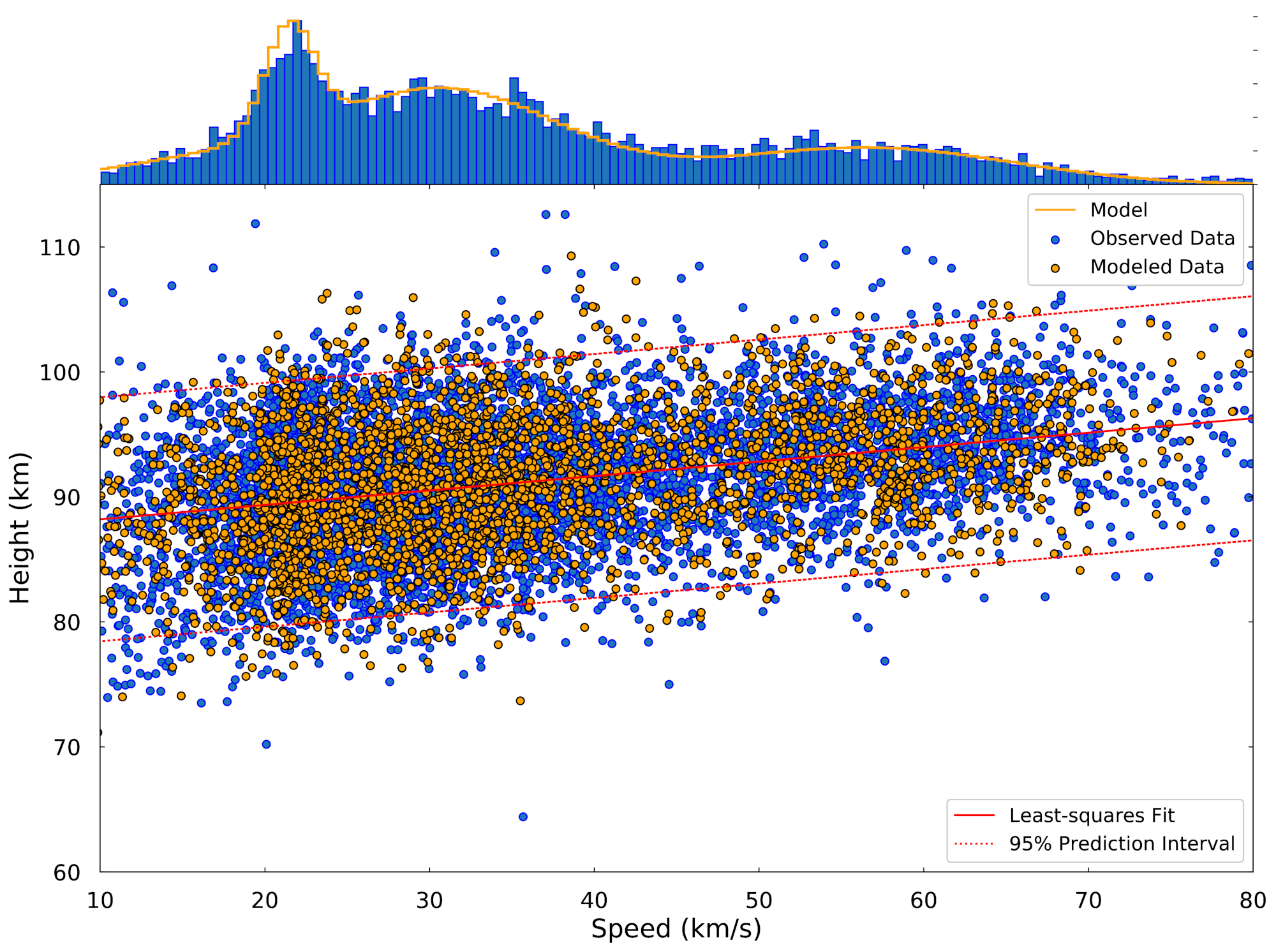}
    \caption{A total of 3000 synthetic echoes generated based on the observed distribution of CMOR echoes from $ \lambda_\odot$=195$\degree$, in 2012. Approximately 6000 observed meteors are shown (blue) with 3000 modeled meteors (orange) generated to match the height vs. speed relation of the observed data (shown as an overlay). The best-fit to the height vs. speed data is shown as the solid line while the 95$\%$ prediction interval is bounded by the dotted lines. The top plot shows the observed \add[mazur]{TOF} speed distribution (histogram bins)\note[Gunter]{observed are TOF speeds??} and the model distribution generated to emulate the measurements (yellow line). }
    \label{fig:ModelDistribution}
\end{figure}

\begin{figure}
    \centering
    \includegraphics[width=\linewidth]{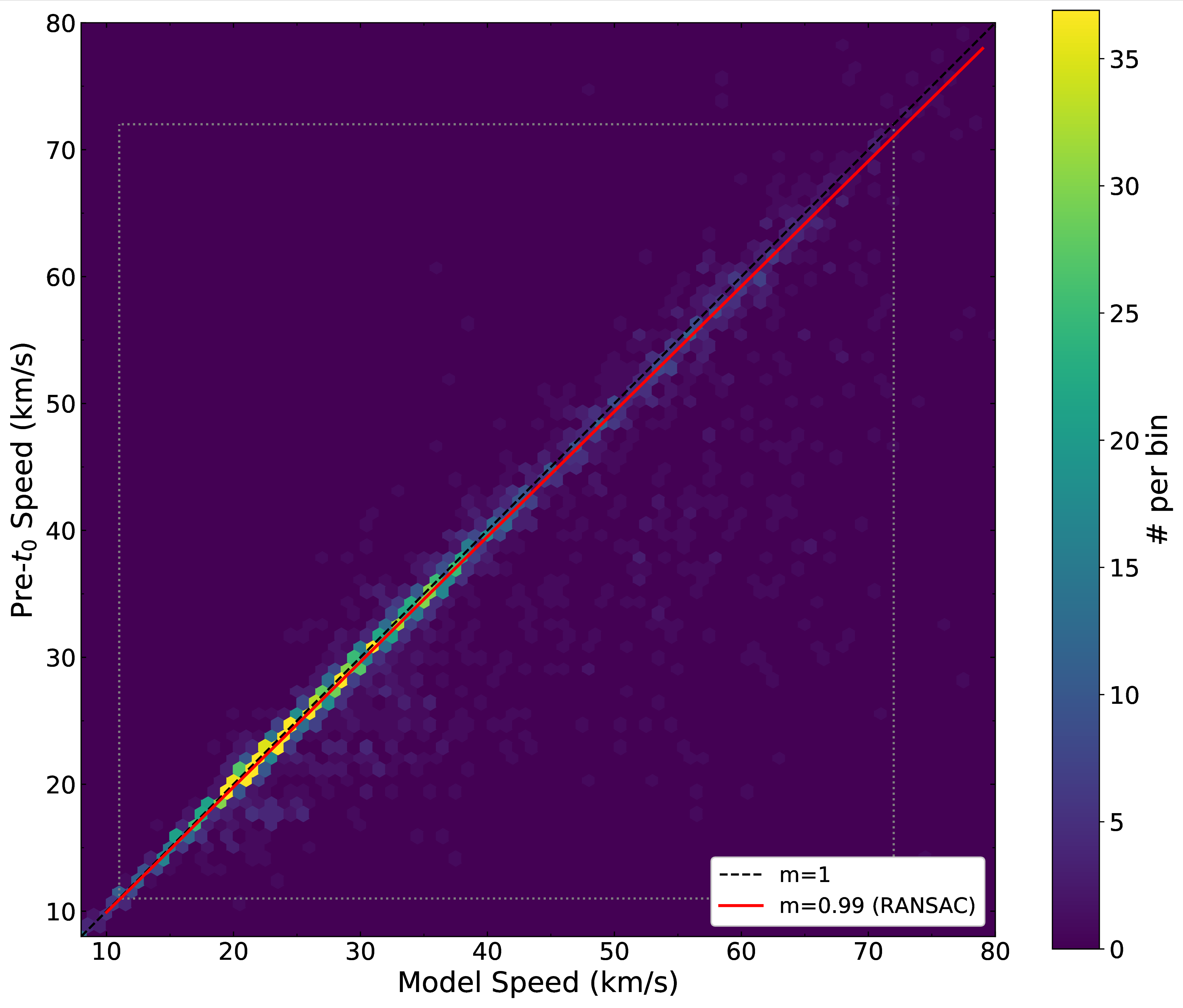}
    \caption{A density plot of pre-$t_{0}$ speeds determined from the PSSST method versus model echo speeds for synthetic data conditioned to match height vs. speed of real echoes detected on Oct 9, 2012. The dotted square shows the 11 km/s and 72 km/s extreme velocity bins where objects are in geocentric orbits (lower) or hyperbolic relative to the sun (upper).}
    \label{fig:ModelHexbin}
\end{figure}

In order to verify the accuracy of the algorithm, 3000 CMOR\remove[R2]{ }-type simulated echoes were generated. Input parameters were generated by a simplistic Monte Carlo simulation using the observed distribution in speed and range/height from CMOR measurements on Oct 9, 2012 to produce equivalent synthetic echoes. Data from this day \remove[R1]{(Oct 9 , 2012)} were chosen as a low speed shower (October Draconids) had a strong outburst and are well represented as a distinct population within the data, a useful feature for comparison. Figure \ref{fig:ModelDistribution} shows the real and modeled distributions for CMOR data from 2012 data at a solar longitude of 195\add[R2]{\degree}. The observed speed distribution, including the Draconid meteor shower peak at about 21 km/s, was modeled as a coarse KDE using multiple Gaussians at 5 discrete speeds. The mean height and standard deviation as a function of speed was defined from the observed echoes on this day as shown in Figure~\ref{fig:ModelHexbin}. Although this particular data\add[R1]{ }set does not represent a typical day (as it captures the Draconid meteor storm of 2012), it serves as a useful test of the modeling method.

These synthetic data were run through the PSSST algorithm and the resultant speeds compared to the known input speeds (Figure \ref{fig:ModelHexbin}). The results show that, although there is some scatter in the data, there is a near 1:1 relation between pre-$t_{0}$ and modeled speeds. When a RANSAC regressor is used with 5 km/s outlier discriminator, the slope is found to be 0.99, demonstrating that PSSST reproduces model speeds very well.

One limitation of the pre-t0 method via PSSST (or any other approach to radar meteor speeds) is the radar sampling rate. If the rate of phase change prior to the $t_{0}$ point exceeds the Nyquist sampling frequency (for CMOR around 266 Hz) the phase unwrapping will alias. In theory, more than two samples are required for robust unwrapping of the phase; more typically, we find that 5 or 6 samples are required for reliable unwrapping and as many as 10 samples may still lead to aliasing in some cases. \add[petr]{For example, requiring six samples per cycle sets the upper aliasing limit for meteor speeds to 84 km/s when we consider echoes with ranges around 100 km and radar wavelengths of 10 m, i.e., the most common detection configuration for CMOR.}

\begin{figure}
    \centering
    \includegraphics[width=\linewidth]{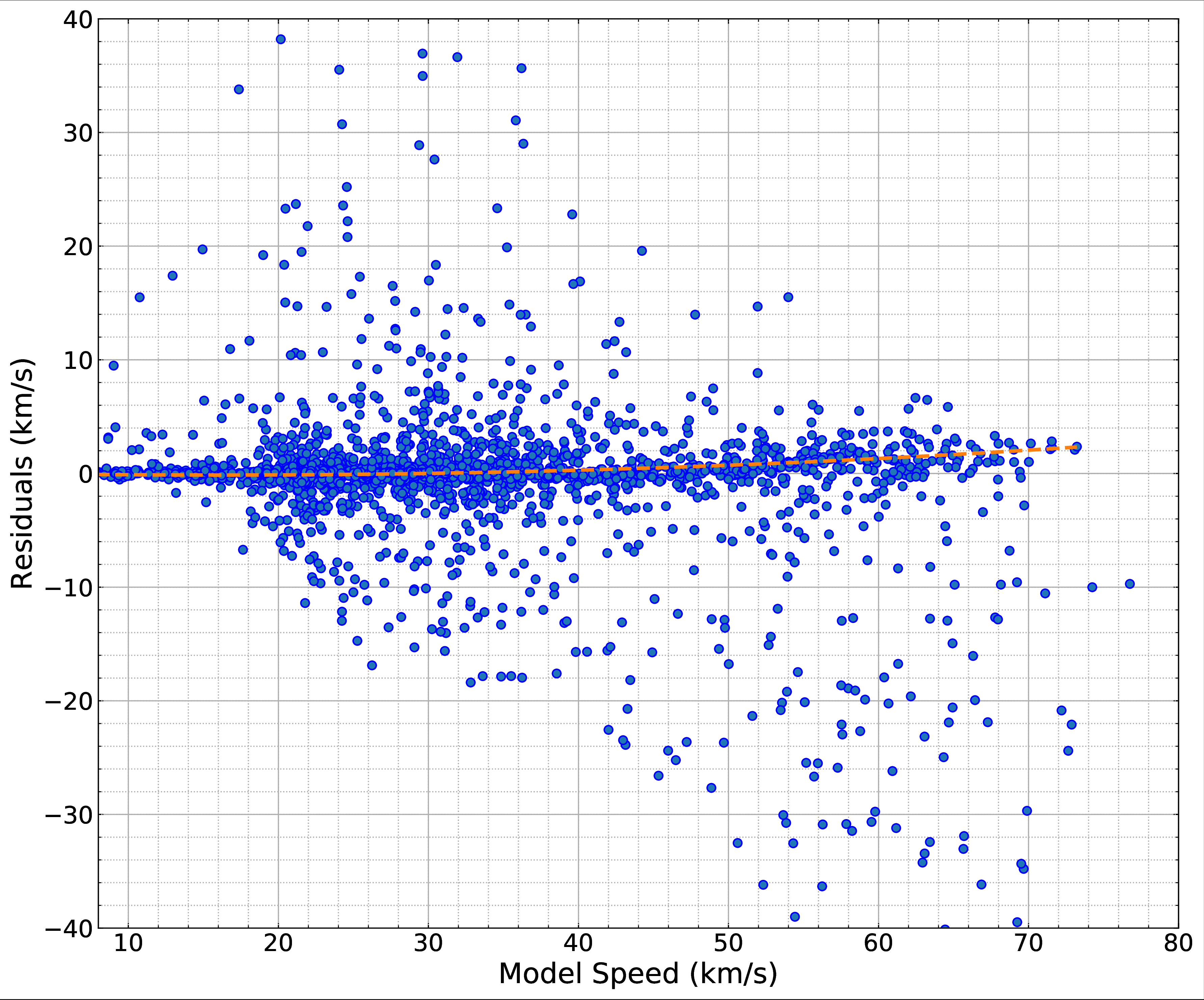}
    \caption{PSSST-\change[R1]{TOF}{Model} residuals for model echoes showing the low-spread of residuals ($\leq \pm 3$ km/s) for speeds less than 20 km/s. Above 20 km/s, the residuals are higher, ranging between about 4-6 km/s.}
    \label{fig:ModelResiduals}
\end{figure}

Figure \ref{fig:ModelHexbin} shows that for speeds below 20 km/s, there is less absolute uncertainty in the pre-$t_{0}$ solution. In general, the standard deviation in the pre-$t_{0}$ residuals is less than about 3 km/s up to a speed of 20 km/s and then varies between 4 and 6 km/s over the rest of the speed range (see Fig. \ref{fig:ModelResiduals}). This can be understood as at lower speeds there is a shallower phase change at any given Fresnel interval prior to the $t_{0}$ point. On the other hand, lower speed echoes of a given mass will have lower SNRs than high speed echoes (ignoring height ceiling effects) - which may reduce precision of the measurement. \add[mazur]{The SNR of high-speed echoes, however, also suffers due to height ceiling effects. Since meteor height increases with speed and atmospheric density decreases with height, the initial radius of a meteor's ionization trail also increases with height due to the increasing atmospheric mean free path. And when the initial radius exceeds 1/4 the radar wavelength, the returned phase loses coherency and the signal is attenuated} \cite{CampbellBrown&Jones2003}.

In overall terms, our application of PSSST to synthetic data shows that the algorithm is robust and self-consistent. Validation of the algorithm requires real-world application to data and meteor speeds estimated by independent techniques, which we examine next. 

\section{Observational Comparisons}
\label{sec:results}

\subsection{MAARSY Transverse Scattering Data}

The MAARSY radar is a narrow-beam, high-power large aperture \change[Gunter]{VHF}{HF} radar, normally used to measure head echoes \cite{SCHULT2017} and dynamics \cite{Stober_2018_maarsy_dynamics}. However, as its main beam is \change[Gunter]{wide}{large} enough to capture most transverse scattering from specular echoes, we can use it to validate the algorithm at very small masses. In this mode, MAARSY is able to record meteor echoes with equivalent radio magnitude of +14 to +16, \change[R1]{or}{with} \change[R1]{masses in the $\mu$g range and even smaller}{the majority of echoes being in the mass range between} 10\textsuperscript{-9} to 10\textsuperscript{-11} kg \cite{SCHULT2019}.

The MAARSY data\add[R1]{ }set examined here was collected from September 16\change[mazur]{th}{$^{th}$} to September 22\change[mazur]{nd}{$^{nd}$}, 2016 and consists solely of transverse scattering echoes. It has been manually cleaned to remove non-specular/head echoes. Some echoes showed low SNR and non-standard (broad) amplitude-time profiles \cite<cf.>[]{Cervera_etal_1997} indicative of fast diffusion. These and some fragmenting echoes, all of which were transverse scattering, were left in the data\add[R1]{ }set. After this filtering, a total of 1696 specular echoes were measured using the PSSST method. Note that the radar experiment was such that specular echoes were collected for only a fraction of the total time, spread evenly over the six days. \note[Gunter]{maybe refer for the specular experiment details again to the Schult et al. paper}

\begin{figure}
    \centering
    \includegraphics[width=\linewidth]{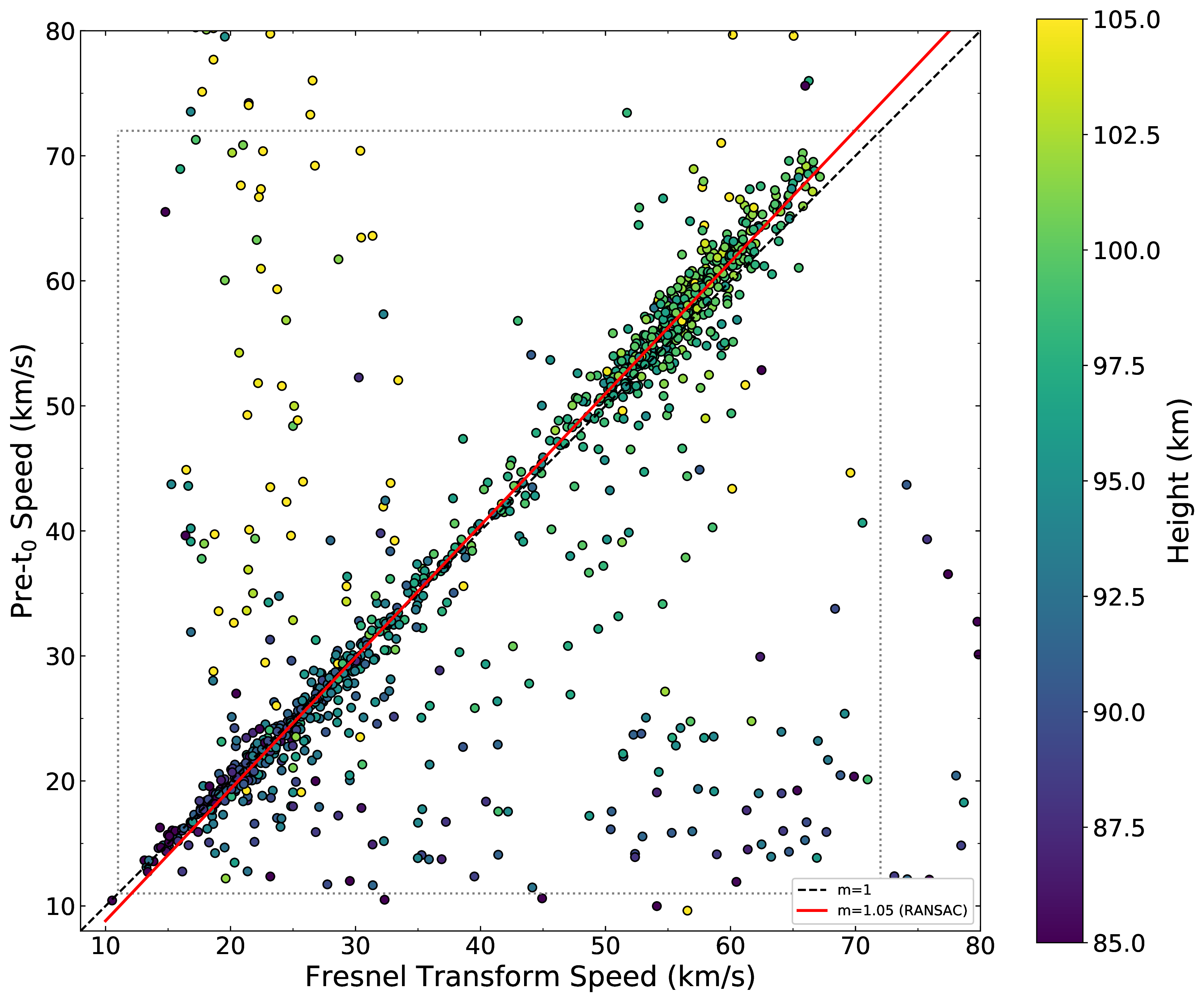}
    \caption{The PSSST measured echo speeds versus FT-based echo speeds for 1530 MAARSY specular echoes coloured by height.}
    \label{fig:maarsy_xplot}
\end{figure}

To compare speeds from PSSST, the data was also processed to estimate a Fresnel transform-based speed implemented as described in \citeA{Holdsworth_etal_2004}.\note[Gunter]{the Fresnel transform paper was 2007???}
Interactive plots of pre-$t_{0}$ vs. Fresnel Transform speeds were examined and the following broad observations made:

\begin{enumerate}
    \item In general, pre-$t_{0}$ and Fresnel Transform (FT) speeds are in good agreement. Although most points plot close to the 1:1 line, it appears that there is good agreement between the methods below speeds of about 40-50 km/s. At higher speeds, the PSSST solution appears to deviate from the FT speeds. A similar trend was seen with the model data, the magnitude of the deviation is greater with the MAARSY data. Presuming this is not due to the FT method, it suggests the PSSST method tends to give velocities that are slightly too high for echoes with speeds greater than about 40 km/s, which would indicate that the $t_{0}$ picks may be somewhat early at higher heights. This is very similar to the behaviour found on \citeA<pg. 183 in>[]{Cervera_1996}, who indicates that high altitude echoes have post-$t_{0}$ phase behaviour which tends to push the point of minimum phase earlier. For PSSST this would directly result in a systematic shift in apparent speed to higher values as observed. 
    \item There is a strong correlation between speed and height, because of the very small masses of MAARSY meteoroids producing specular echoes of $10^{-8}<m<10^{-12}$ kg. Here the mass range was computed based on the received echo power, range and using the mass-electron-line density relation of \citeA{Verniani1973}. This produces a strong height filter and is therefore a direct proxy for speed. This can be used to identify echoes for which one method performs better than the other. In general, most of the echoes that plot in the upper left of Figure \ref{fig:maarsy_xplot} have an observed height greater than 100 km\remove[R2]{.} suggesting that the calculated PSSST speeds may be correct while the FT speed is much too low. Similarly, most of the echoes in the lower right of Figure \ref{fig:maarsy_xplot} occur at heights below about 95 km - also more consistent with the PSSST speeds than the FT speeds.
\end{enumerate}

Of the 1695 processed echoes, 166 (10$\%$) did not produce a PSSST solution due to poor pre-$t_{0}$ phase behaviour. Of the successful echoes, 75$\%$ were within $\pm$10$\%$ of the FT speed. Although FT can be a robust method for calculating meteor speeds, it is not necessarily ground-truth in all cases (nor is PSSST). 

\subsection{CMOR Transverse scattering Data}

\begin{figure}
    \centering
    \includegraphics[width=\linewidth]{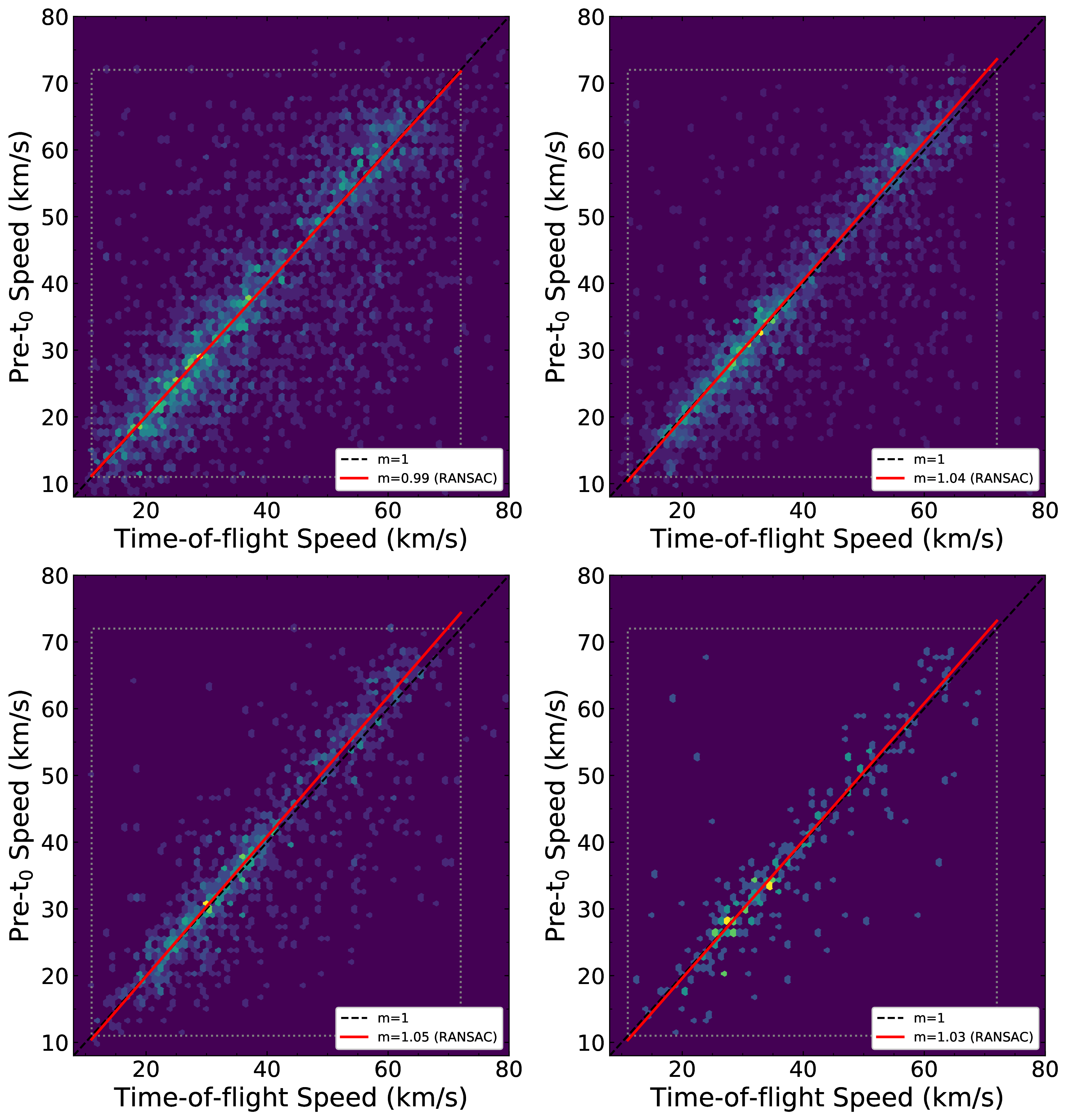}
    \caption{PSSST speeds as a function of TOF speeds for 2018 $ \lambda_\odot$=156$\degree$ grouped according to 3-station (upper left), 4-station (upper right), 5-station (lower left), and 6-station (lower right) TOF solutions.}
    \label{fig:cmor_hexbin-SPOR}
\end{figure}

As a first validation comparison for PSSST using data from a low-power broad-beamed meteor radar, a random single day of meteor echoes measured by CMOR was chosen (from 2018 at a solar longitude ($\lambda_\odot = 156^\circ$)) and echoes grouped according to the number of receiving stations used in the time-of-flight (TOF) solution. Of the 6359 total echoes, there were 2918 3-station echoes, 2111 4-station echoes, 1071 5-station echoes, and 259 6-station echoes (Figure \ref{fig:cmor_hexbin-SPOR}). 

\begin{figure}
    \centering
    \includegraphics[width=\linewidth]{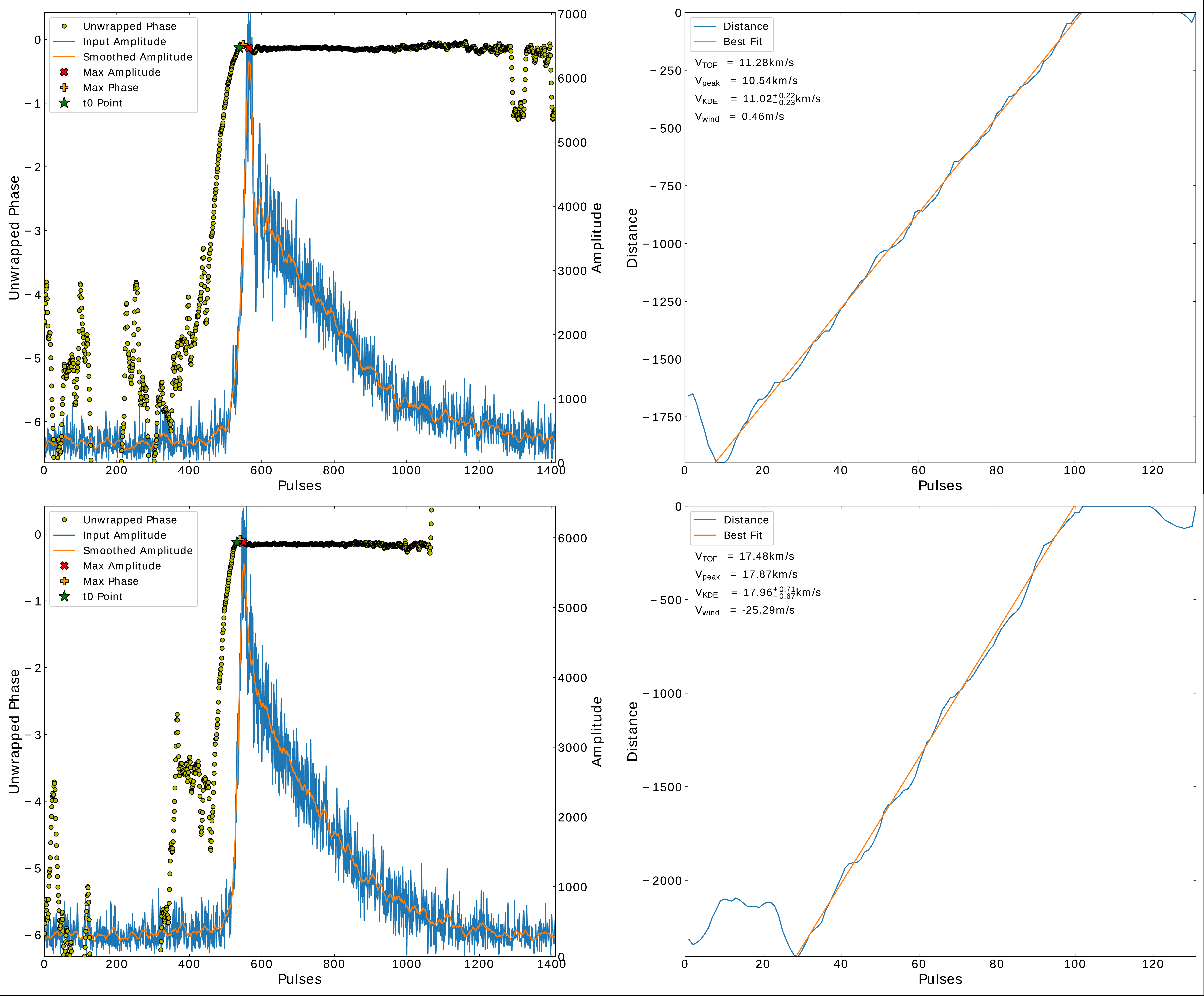}
    \caption{PSSST speed results for echo\remove[R1]{e} examples having TOF speeds of 11.28$\pm$N/A km/s (top panels) and 17.48$\pm$1.12 km/s (bottom panels). The radial neutral wind drift is also shown ($V_\mathrm{wind}$). }
    \label{fig:pssst_results_11-17}
\end{figure}

\begin{figure}
    \centering
    \includegraphics[width=\linewidth]{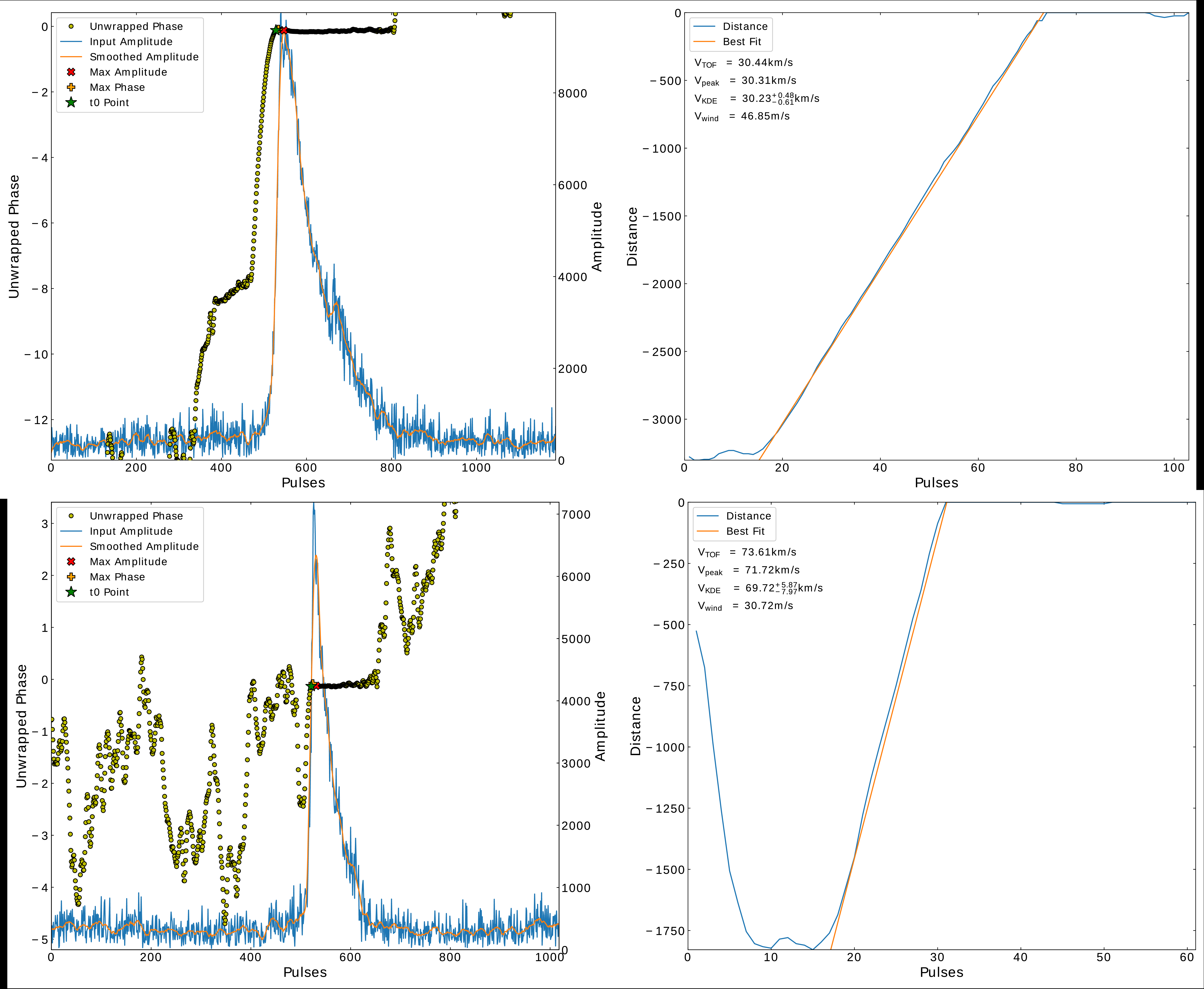}
    \caption{PSSST results for echoes with TOF speeds of 30.44$\pm$0.44 km/s (top panels) and 73.61$\pm$471.4 km/s (bottom panels).\add[R1]{ The large error associated with the TOF solution for the bottom echo is a sign that the TOF algorithm has failed due to poor specular point geometry where all points are near the same point on the trail.}}
    \label{fig:pssst_results_30-73}
\end{figure}

PSSST was applied to each group of echoes in turn. Figures \ref{fig:pssst_results_11-17} and \ref{fig:pssst_results_30-73} show examples of echoes of different speeds. In each case, the signals are well-behaved and the resultant solutions are similar to the TOF solutions. The final example\add[R1]{ (lower panels, Figure }\ref{fig:pssst_results_30-73}), however, with a TOF speed of 73.61 $\pm$471.4 km/s produces a PSSST speed that is much lower, at $69.72^{+5.9}_{-8.0}$ km/s. \add[R1]{The uncertainty in the automated TOF solution is orders of magnitude larger than what we would expect, and is the first sign that the TOF method has failed. In this case, the TOF failure is due to the poor specular point geometry between the stations - ie. they all have specular points at nearly the same location on the trail. Since the echo has a high SNR, manual TOF processing would likely give a much better solution, though still with large error. }Given that the best-fit \add[R1]{PSSST }solution closely matches the slope of the distance-pulse curve and represents a non-hyperbolic orbit with a reasonable uncertainty range, it is likely that the PSSST solution is better than the TOF solution in this particular case. In fact, if we make the simple assumption that hyperbolic orbits are those with speeds greater than 72 km/s\add[R1]{ (the upper limit for a bound Solar System object to impact Earth)}, we find that PSSST returns fewer apparent hyperbolic solutions than does the TOF method. Of the more than 12.4 million echoes processed from 2018, 1.9$\%$ had TOF speeds greater than 72 km/s while only 0.6$\%$ of the PSSST speeds were greater than 72 km/s.

\begin{figure}
    \centering
    \includegraphics[width=\linewidth]{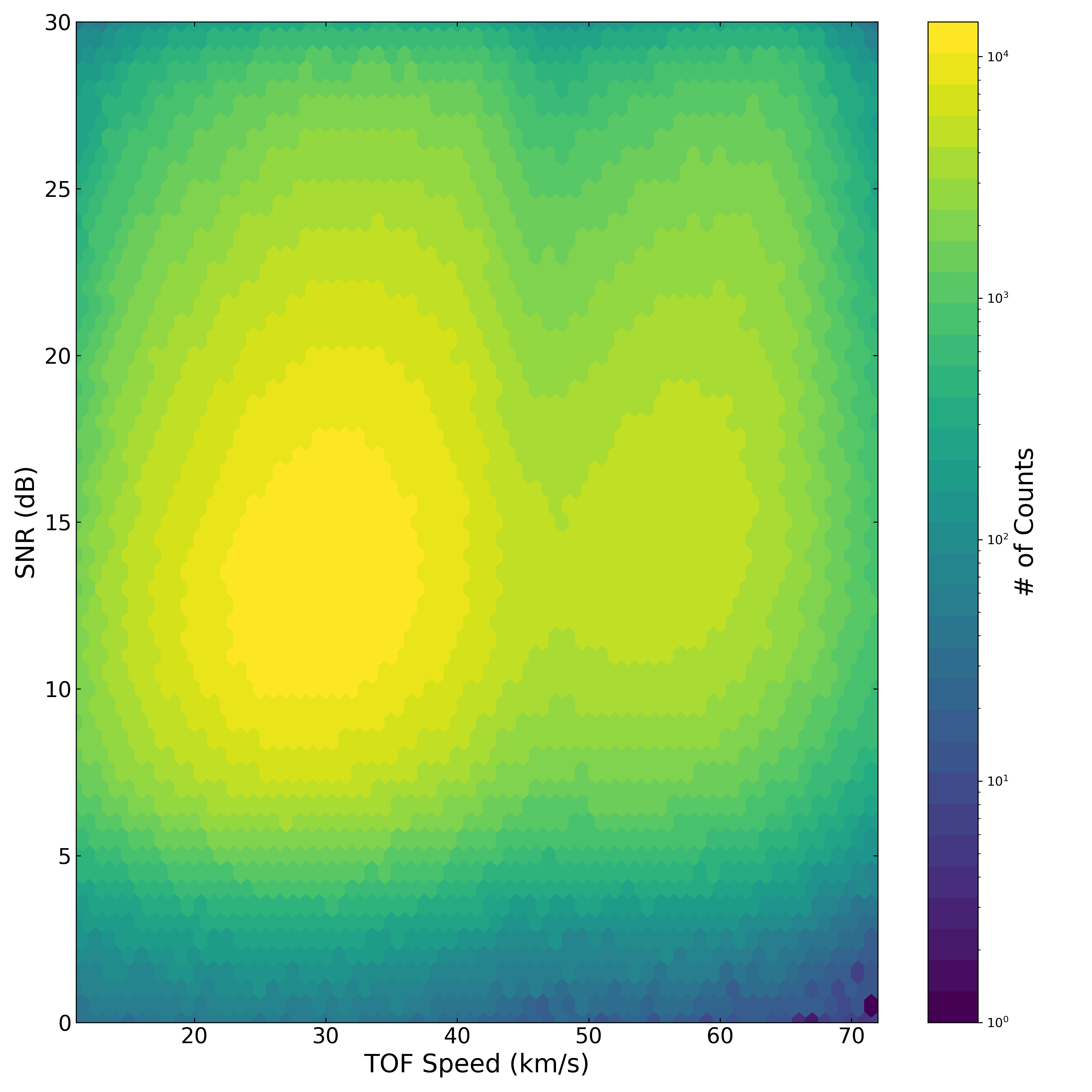}
    \caption{The overall \add[mazur]{log} distribution of signal-to-noise ratio of 12 million CMOR echoes where TOF speeds were measurable from data collected between 2010-2019.The broadening in the distribution at higher speeds shows that the proportion of high-speed echoes in the data set increases with increasing SNR.}
    \label{fig:SNRvsSpeed}
\end{figure}

\begin{figure}
    \centering
    \includegraphics[width=\linewidth]{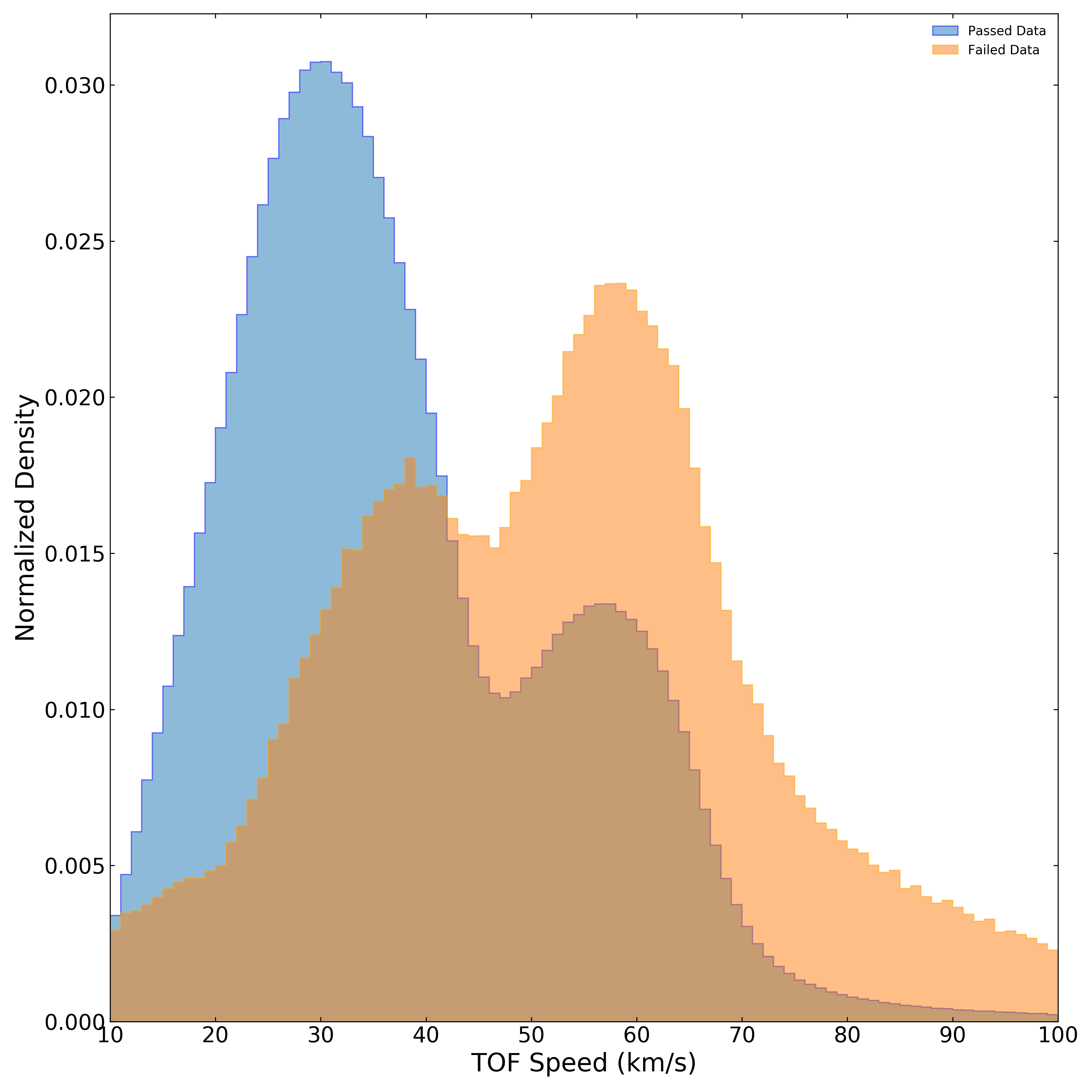}
    \caption{The relative fraction of echoes per one km/s bin as a function of TOF speed showing where the PSSST algorithm produced a speed estimate (passed) and failed to converge to a speed estimate (failed).}
    \label{fig:TOFdensity}
\end{figure}

To examine the effect of SNR on the PSSST success rate, the SNR of CMOR echoes having time-of-flight speed measured from nine years of data (2010-2019) are plotted in Figure \ref{fig:SNRvsSpeed}. At lower speeds there is a strong peak in the number of echoes with SNRs in the range of about 12-16\add[R2]{ dB}, while at higher speeds, the echo SNRs are skewed slightly higher with \change[mazur]{no obvious}{a broader, less-defined} peak. This shows that the general proportion of high-speed echoes in the total data\add[R1]{ }set increases with increasing SNR. This trend in TOF SNRs implies that the PSSST failure rate should also increase with increasing SNR. This is simply because of the increased chance of high-speed echoes showing aliasing in phase \add[mazur]{and }not permitting good PSSST measurements. As an illustration of this increasing number of failures with speed, Figure \ref{fig:TOFdensity} shows a density plot of the data arranged into passed (where PSSST could resolve a speed) and failed (where the algorithm failed to converge to a speed) data.

\begin{figure}
    \centering
    \includegraphics[width=\linewidth]{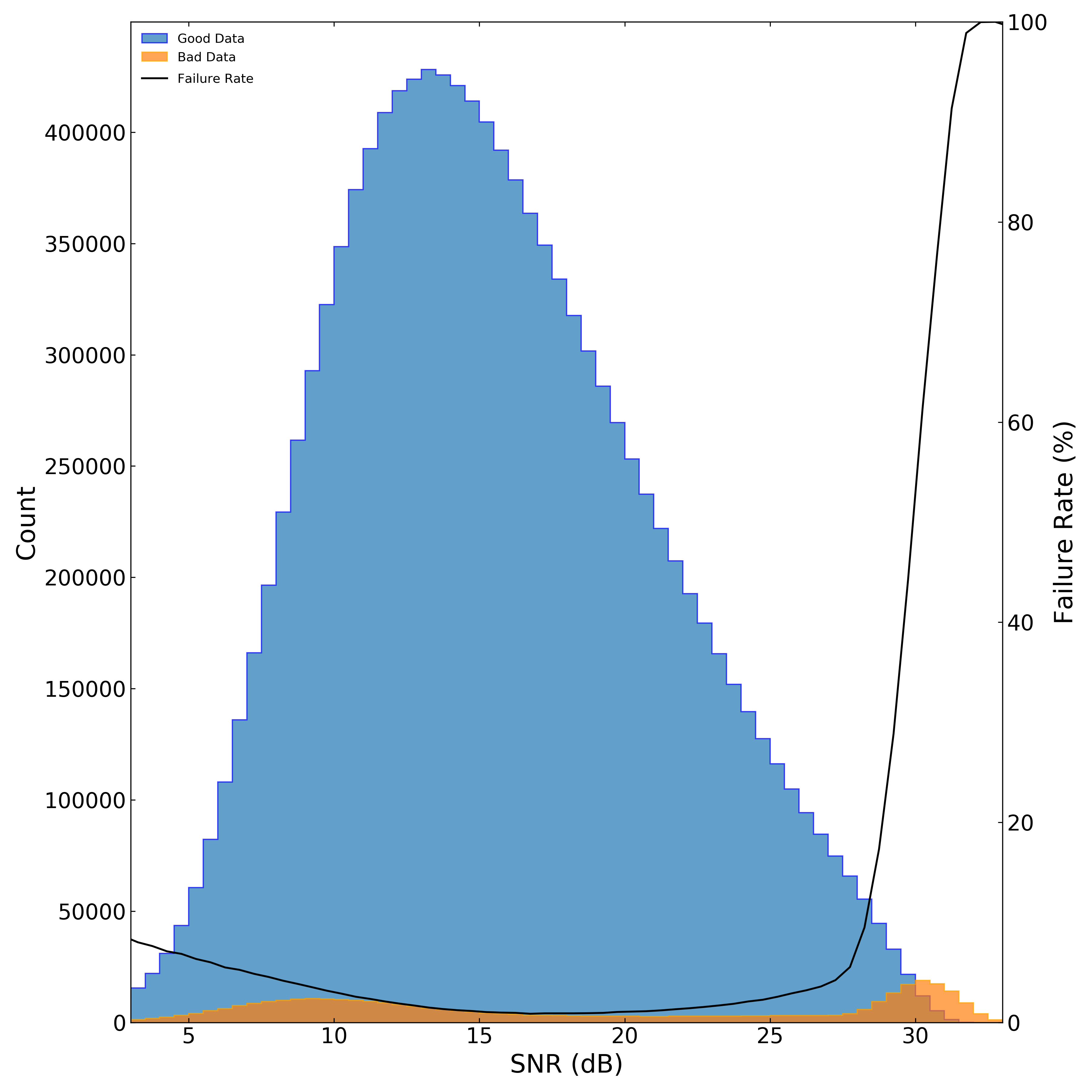}
    \caption{Shown are a CMOR data\add[R1]{ }set of 12 million echoes with data considered to be good if both pre-$t_{0}$ and TOF solutions exist and the measured amplitude is less than 25,000 DU \note[Gunter]{DU-digitizer units}. The distribution of good data peaks at an SNR of about 17, while the bad data peaks at an SNR of about 9.5\add[R2]{ dB}. The failure rate is generally high for low SNR echoes and low for SNRs in the middle of the range. At higher SNRs, failures increase due to the applied amplitude filter and increasing number of higher speed echoes at higher SNR that may fail due to aliasing effects. }
    \label{fig:SNRhistogram}
\end{figure}

Figure \ref{fig:SNRhistogram} arranges the solutions into \change[R2]{'good'}{``good''} and \change[R2]{'bad'}{``bad''} data sets plotted as a histogram. Here an echo is considered to be \change[R2]{'good'}{``good''} when both TOF and pre-$t_{0}$ solutions exist and when the maximum amplitude of the echo is less than 25,000 \note[Gunter]{DU-digitizer units}. This latter criterion ensures that over-saturated echoes are not being included in the \change[R2]{'good'}{``good''} data set. All other data is then considered to be \change[R2]{'bad'}{``bad''}. The histogram in figure \ref{fig:SNRhistogram} shows that the \change[R2]{'good'}{``good''} data peaks at an SNR of about 13\add[R2]{ dB}, while the \change[R2]{'bad'}{``bad''} data peaks at an SNR of about 9.5\add[R2]{ dB}. From this, the overlaid curve shows that the failure rate reaches about 8$\%$ at low SNRs and is at a minimum of about 1$\%$ at an SNR of 17\add[R2]{ dB}. For higher SNRs, the failure rate appears to increase, a result of the maximum amplitude filter that was applied as well as the increasing proportion of higher speed echoes at higher SNRs that are more likely to have a failed PSSST solution due to aliasing. On average, the failure rate is 2.9$\%$ when the high amplitude filter is present and 2.1$\%$ when only pre-$t_{0}$ failures are considered to be bad echoes.

As expected, the SNR of the echo is the greatest predictor of the success of the method. Poor-quality (noisy) phase data can result in errors greater than 10's of km per second or complete failure to compute a speed. Unfortunately, low-SNR phase data typically has even poorer quality amplitude data. This is expressed as an increase in failure rate of computed speeds at low SNR which makes accurate speed determination difficult by any method. To mitigate this, it would be useful to first apply a filter based on SNR before using the PSSST. For CMOR data, we can accomplish this by filtering either directly by SNR or indirectly by the number of stations available for a TOF solution\remove[R2]{(N$_\mathrm{TOF}$)}.

\begin{figure}
    \centering
    \includegraphics[width=\linewidth]{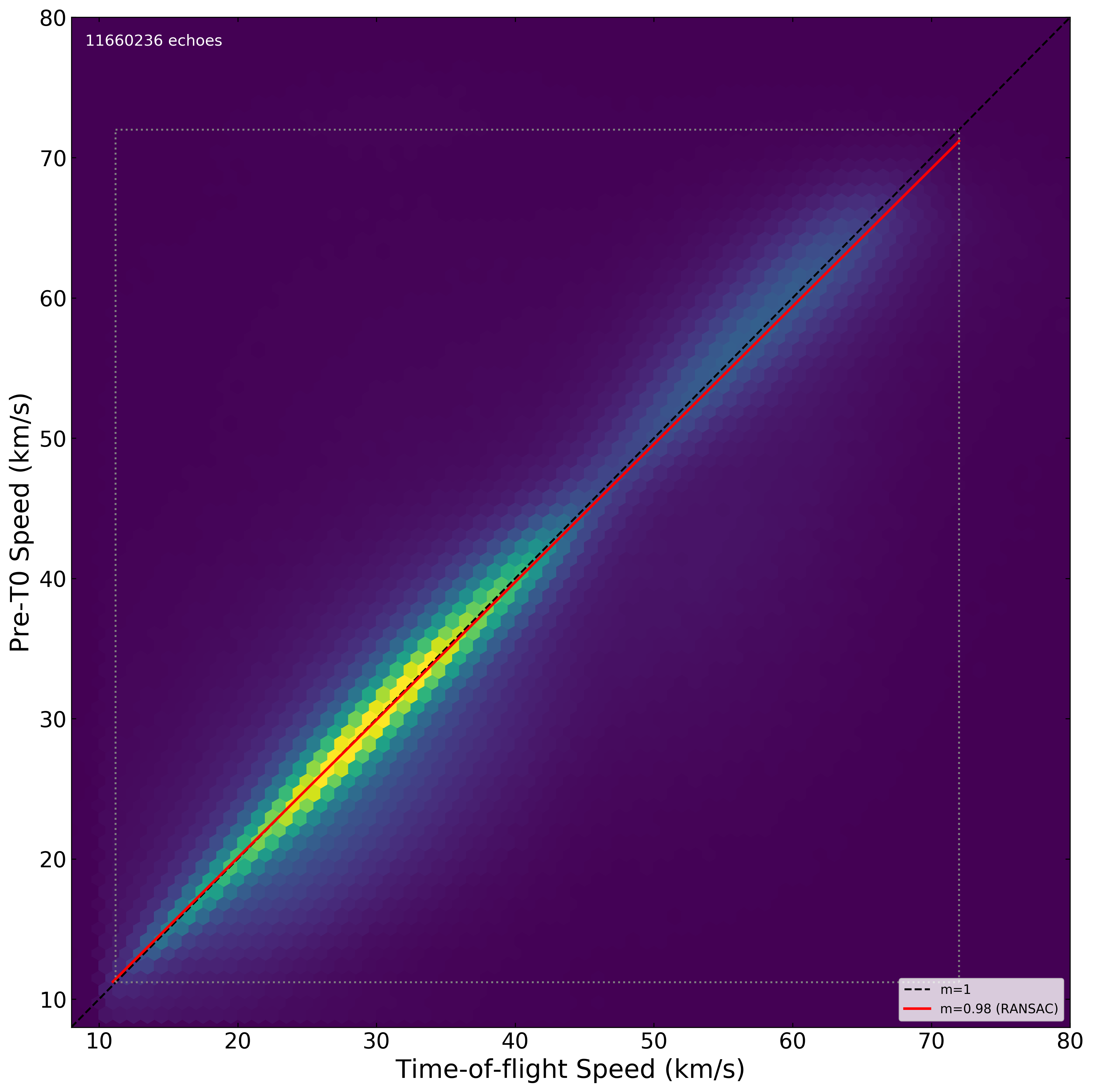}
    \caption{A heat plot of PSSST speeds \change[R1]{vs.}{versus} TOF speeds for more than 11.6 million CMOR echoes.}
    \label{fig:hexbinallyears}
\end{figure}

\begin{figure}
    \centering
    \includegraphics[width=\linewidth]{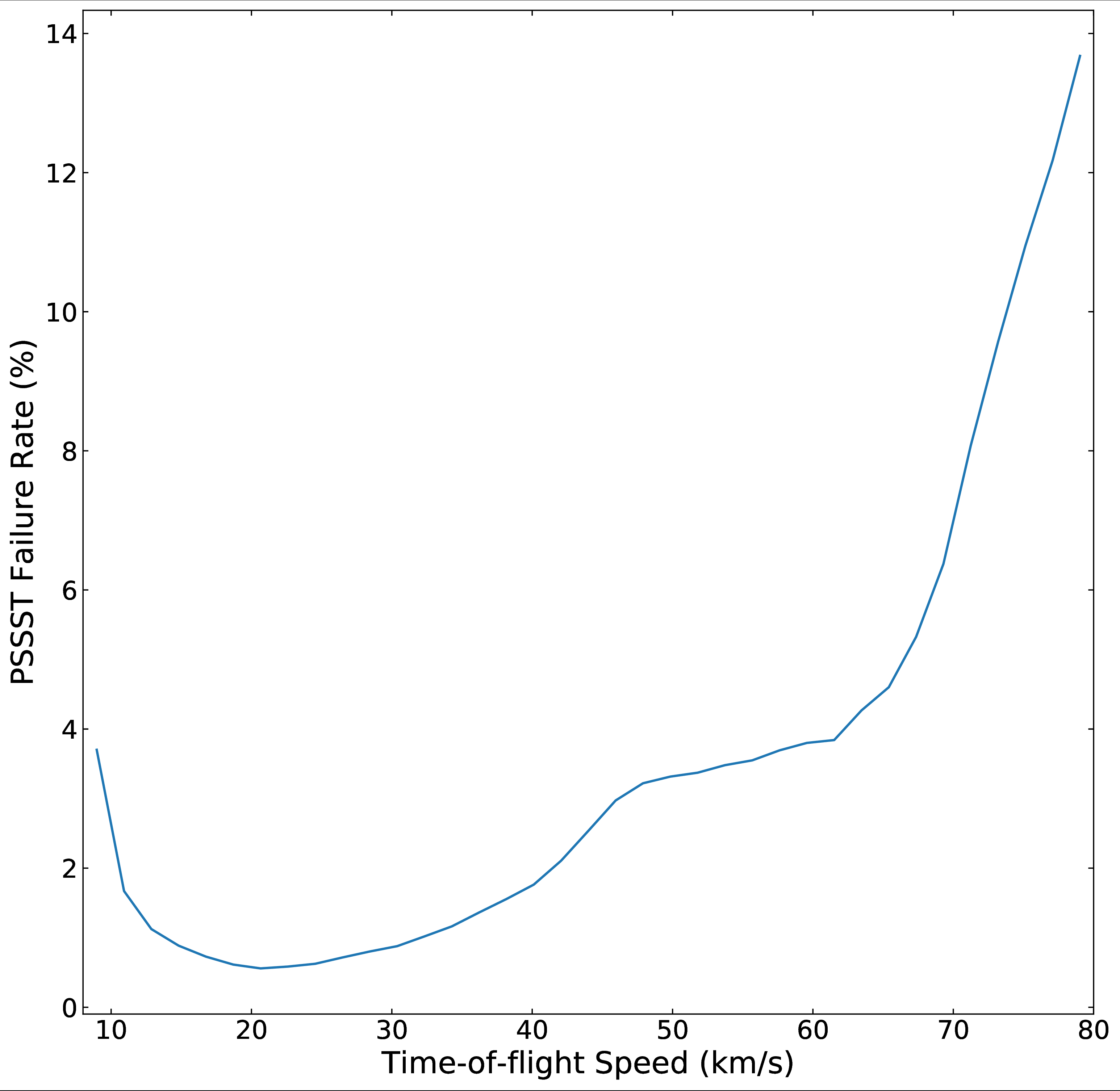}
    \caption{The PSSST failure rate as a function of meteor speed  for CMOR 3+-station data. The failure rate ranges from less than 1$\%$ to 14$\%$.}
    \label{fig:failurerate}
\end{figure}

\begin{figure}
    \centering
    \includegraphics[width=\linewidth]{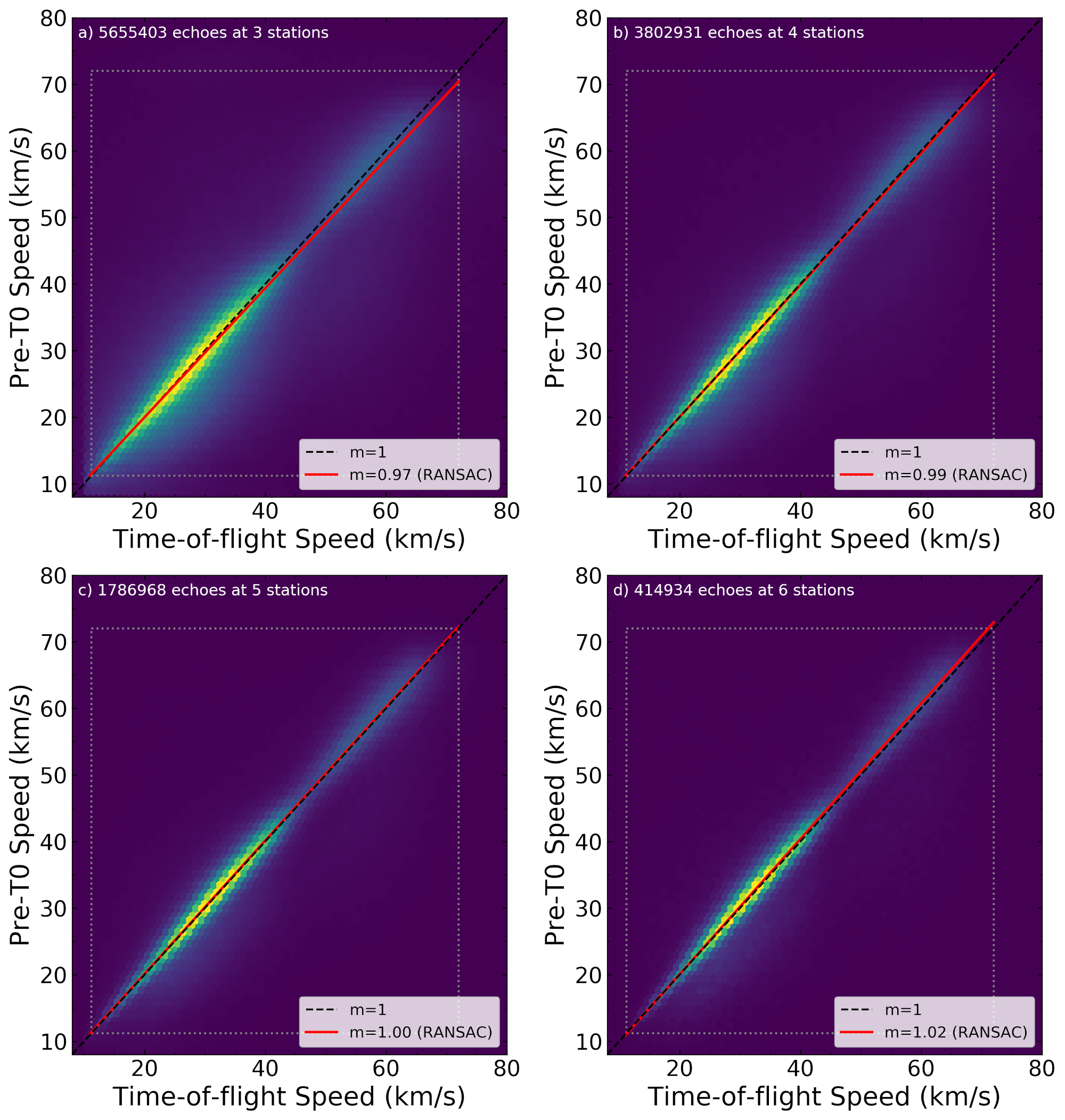}
    \caption{Plots of more than 11.6 million echoes grouped by number of receiving stations.}
    \label{fig:hexbinmultiallyears}
\end{figure}

In general, we find that the PSSST speeds are well correlated with the TOF speeds (Figure \ref{fig:hexbinallyears}). The roughly 1:1 correlation does not appear to be affected by the number of stations used in the TOF solutions. The scatter about the 1:1 line, however, does increase with decreasing station count - presumably due to noisier echoes and decreasing TOF speed precision for echoes with lower station counts. As with the MAARSY data, processing failures were typically due to poor pre-$t_{0}$ phase information. Failure rates were determined to be 2$\%$, 1$\%$, and $<1\%$ for all, 3-station, 4-station, 5/6-station echoes respectively. Using the entire data\add[R1]{ }set, the failure rate (Figure \ref{fig:failurerate}) is speed dependent and ranges from less than 1$\%$ to nearly 14$\%$ with an average across an 8-80 km/s range of 2.2$\%$. Again, when the data is examined on the basis of number of receiving stations, the \change[R2]{correlation}{slope of the best-fit line} between PSSST and TOF speeds is nearly 1:1 (Figure \ref{fig:hexbinmultiallyears}).

Although these rates are low when compared to other methods, it must be kept in mind that the lowest SNR echoes have been filtered out on the basis of the 3-station-minimum requirement for the TOF solutions. These \change[R1]{multi-}{$<$3 }station echoes represent only 48$\%$ of the echoes observed on this day and we expect \change[R1]{the }{an increase in PSSST }failure rate \change[R1]{will increase when these $<$3 station echoes are added to the total}{if they are included}.\add[R1]{ However, we chose to not include them in our analysis as, although complete failure to compute a PSSST solution would be obvious, no TOF solutions would be available for these echoes to measure the accuracy of 'successful' PSSST solutions.}

\subsection{Comparison with shower data}
\begin{figure}
    \centering
    \includegraphics[width=\linewidth]{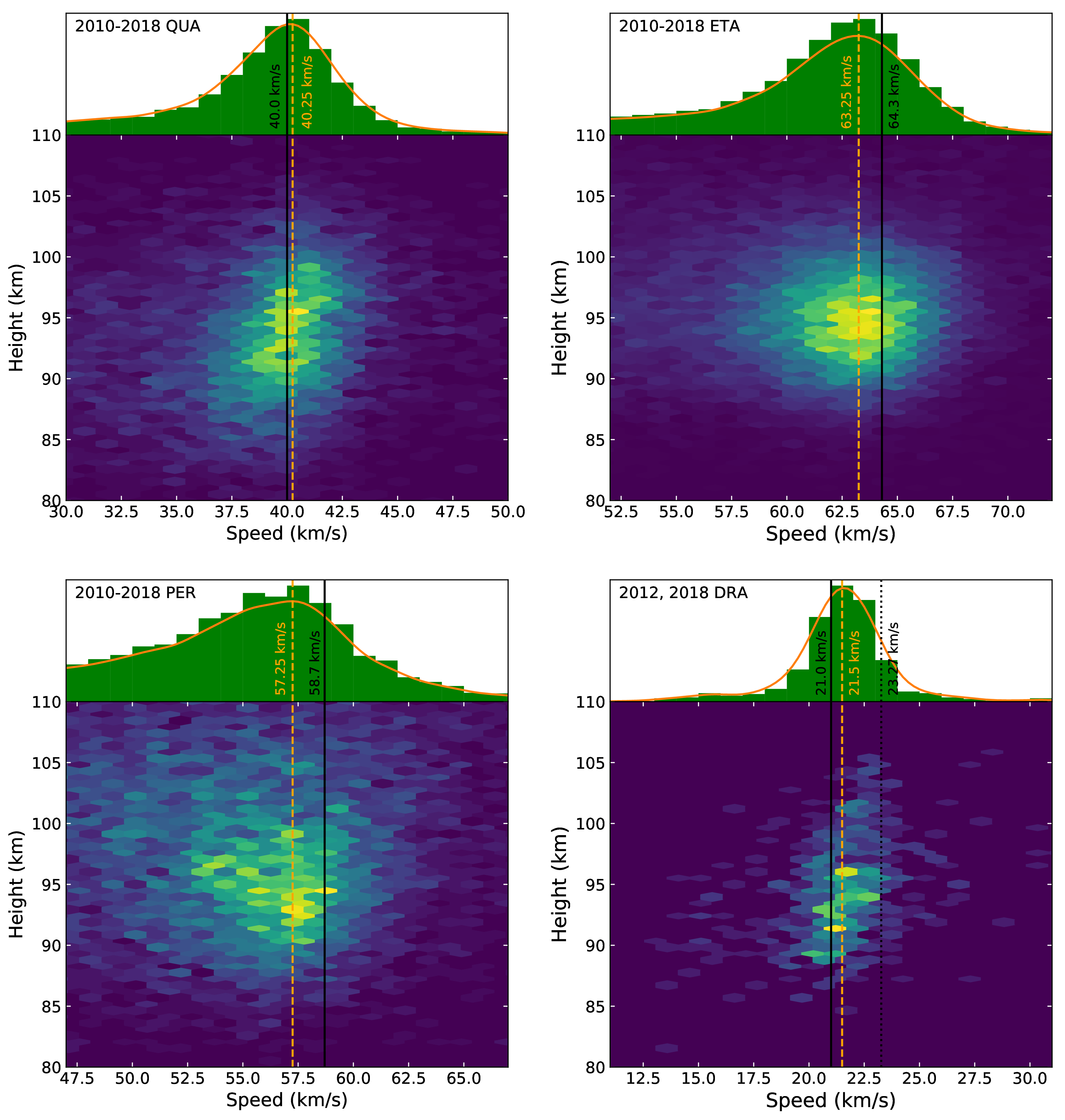}
    \caption{Height-speed distributions for four meteor showers measured using PSSST.}
    \label{fig:ShowerHeightSpeed}
\end{figure}

As a final validation of the PSSST\remove[R2]{ technique}, CMOR meteors from the 2018 Eta Aquarid (ETA), Perseid (PER), Quadrantid (QUA), and 2012 Draconid (DRA) showers (Table \ref{tab:CMORShowerSpecs}) were processed using the PSSST algorithm.\add[R2]{ The showers consisted of between 542 (Draconids) and 36,137 (Eta Aquarids) meteors which were selected from the CMOR data base based on solar longitude ($\lambda_\odot$), velocity, and radiant position filters.} We take the most probable pre-atmospheric speeds for these showers to be those found by \citeA{Schult_etal_2018} from MAARSY head echo measurements, as head echoes are recorded by MAARSY at comparatively high altitudes (Figure \ref{fig:ShowerHeightSpeed}) and hence suffer little from atmospheric deceleration. For the Draconids, we examine speeds found by \citeA{Maslov_2011} and \citeA{Jenniskens_etal_2011} as well as the most probable pre-atmospheric, deceleration-corrected, $v_\infty$ speed from \citeA{Ye_etal_2013}.

The upper panel for each shower in Figure \ref{fig:ShowerHeightSpeed} shows the histogram of the computed speeds (bars) with a KDE overlay (curve). As expected, the mean PSSST speeds and KDE profiles are skewed to lower values than the MAARSY speeds, since the former are uncorrected for deceleration. 

In general, there is good correlation between mean shower PSSST speeds and MAARSY initial speeds for both the Quadrantids (40.25 km/s vs. 40.0 km/s) and the Eta Aquarids (63.5 km/s vs. 64.3 km/s). These 0.6\% and 1.2\% differences for the Quadrantids and the Eta Aquarids are small and reflect the comparatively small \change[mazur]{decelerations}{deceleration} experienced by CMOR-sized meteoroids at these high speeds and heights. 

For the Draconids, the PSSST speeds are distributed around 21.5 km/s which is close to the 21.0 km/s and 20.9 km/s found by \citeA{Maslov_2011} and \citeA{Jenniskens_etal_2011} respectively. Since the Draconids tend to be fragile, they will decelerate more quickly than more compact and less fragile meteoroids. The work of \citeA{Ye_etal_2013} corrects for this to give the pre-atmospheric, deceleration-corrected, $v_\infty$ speed (23.27 km/s) shown by the dotted line on the Draconid panel in Figure \ref{fig:ShowerHeightSpeed}. The deceleration-corrected speed will always be higher than the peak of the distribution of observed speeds and the trend should become asymptotic with $v_\infty$ as height increases. Even with the relatively low number of Draconids observed, we can see that this is the case. 

However for the Perseids, the PSSST speeds tend to be clustered around lower speeds than would be expected from the MAARSY (PER - 57.0 km/s \change[R1]{vs.}{versus} 58.7 km/s). Although this is larger (2.9\%) and seems unacceptable on the surface, there is a likely explanation. A large scatter in the height-speed plot for this shower likely has several sources. The main one being that the Perseids are \change[R1]{near a}{within the North Apex} sporadic source \change[mazur]{with a}{which has} similar speed\add[mazur]{s}, \change[mazur]{which complicates}{thereby complicating} the interpretation. That being said, the TOF speeds for this same data show a similar scatter, so we suspect this represents a system-wide bias rather than a specific failure of the PSSST algorithm. 

\begin{table}
    \caption{Details of selected showers used for validation of the PSSST method$^{a}$}
    \centering
    \begin{tabular}{l c c c c}
    \hline
    Shower &  $\lambda_\odot$ & \# of Meteors & Literature Speed  & PSSST Speed \\
    &&& [km s$^{-1}$] & [km s$^{-1}$] \\
    \hline
    2010-2018 Quadrantids & 281-285 & 8819 & 40.0$^1$ & 40.25\\
    2010-2018 Eta Aquarids & 30-68 & 36137 & 64.3$^1$ & 63.25\\
    2010-2018 Perseids & 124-146 & 10643 & 58.7$^1$ & 57.0\\
    2012, 2018 Draconids & 195 & 542 & 21.0$^2$ & 21.5\\
    \hline
    \multicolumn{5}{l}{$^{a}$Literature values for the shower speeds from $^1$\citeA{Schult_etal_2018} and $^2$\citeA{Maslov_2011}.}
    \end{tabular}
    \label{tab:CMORShowerSpecs}
\end{table}

\section{Conclusions}
In this work, we describe a robust method for calculating pre-$t_{0}$ meteor echo speeds based on the theory of \citeA{Cervera_etal_1997}. This technique has been implemented and applied to CMOR and MAARSY specular echoes. It allows for the automated calculation of speeds and serves as an effective confidence metric when compared to time-of-flight speeds.

The use of a scanning and expanding window for slope determination eliminates problems associated with linear regression over large pre-$t_{0}$ windows and provides automated solutions requiring little in the way of manual quality control. The method incorporates an automated $t_{0}$ picking algorithm that is robust and whose uncertainties are quantifiable by interpretation of the shape of the PSSST speed KDE. Model CMOR echoes indicate that, for speeds less than 50 km/s, $t_0$ picking errors of up to $\pm$6 pulses \change[R1]{can}{will, for most meteors,} still give results within 5$\%$ of the true speed. At higher speeds, some of this robustness is lost with a $t_0$ pick requirement of $\pm$2 pulses to achieve a 5$\%$ accuracy in speed.

Validation of the PSSST method has been done by comparing the computed speeds of meteors from four showers to the speeds for those same showers reported by \citeA{Schult_etal_2018} and \citeA{Ye_etal_2013}. The PSSST results are in good agreement with only minor differences - which are likely a result of deceleration effects and contamination by sporadic\change[R1]{s}{ meteors}.

In addition to shower speed validation, PSSST results have been compared to more than 11.6 million time-of-flight solutions from CMOR and nearly 1700 Fresnel transform solutions for MAARSY specular echoes. These results show good correlation with both TOF and Fresnel transform solutions. Failure rates for the MAARSY data are about 10$\%$ after filtering out non-specular events. For multi-station CMOR data, the average failure rate is only about 2$\%$ with a speed-dependent range of failure rates between $<1\%$ and 14$\%$.

Implementation of this algorithm improves the CMOR workflow in a number of ways. First, it provides an independent speed measurement that can be compared to the time-of-flight speeds. This is especially relevant for low-speed meteors which\change[R1]{ tend to}{, because of their low ionization probability, }have low SNR echoes. Second, it provides an analytical determination of the $t_{0}$ point - something which is not done by the time-of-flight algorithm. 

\section{Acknowledgements}
This work was supported in part by the NASA Meteoroid Environment Office under cooperative agreement 80NSSC18M0046. PGB also acknowledges funding support from the Natural Sciences and Engineering Research council of Canada and the Canada Research Chairs program. The authors thank ATRAD \remove[R2]{Pty} for providing the Fresnel Transform software used in this analysis. \add[Gunter]{The data collection and analysis of MAARSY was supported by grant STO 1053/1-1 (AHEAD) of the Deutsche Forschungsgemeinschaft (DFG).}

We thank Dr. Jack Baggaley and an anonymous reviewer for helpful comments which greatly improved an earlier version of this manuscript.

The code and data for testing the performance of our PSSST method has been made available through Mendeley Data (http://dx.doi.org/10.17632/gbpgfskx27.1).

\appendix 
\section{CMOR TOF Solutions}
\begin{figure}
    \centering
    \includegraphics[width=\linewidth]{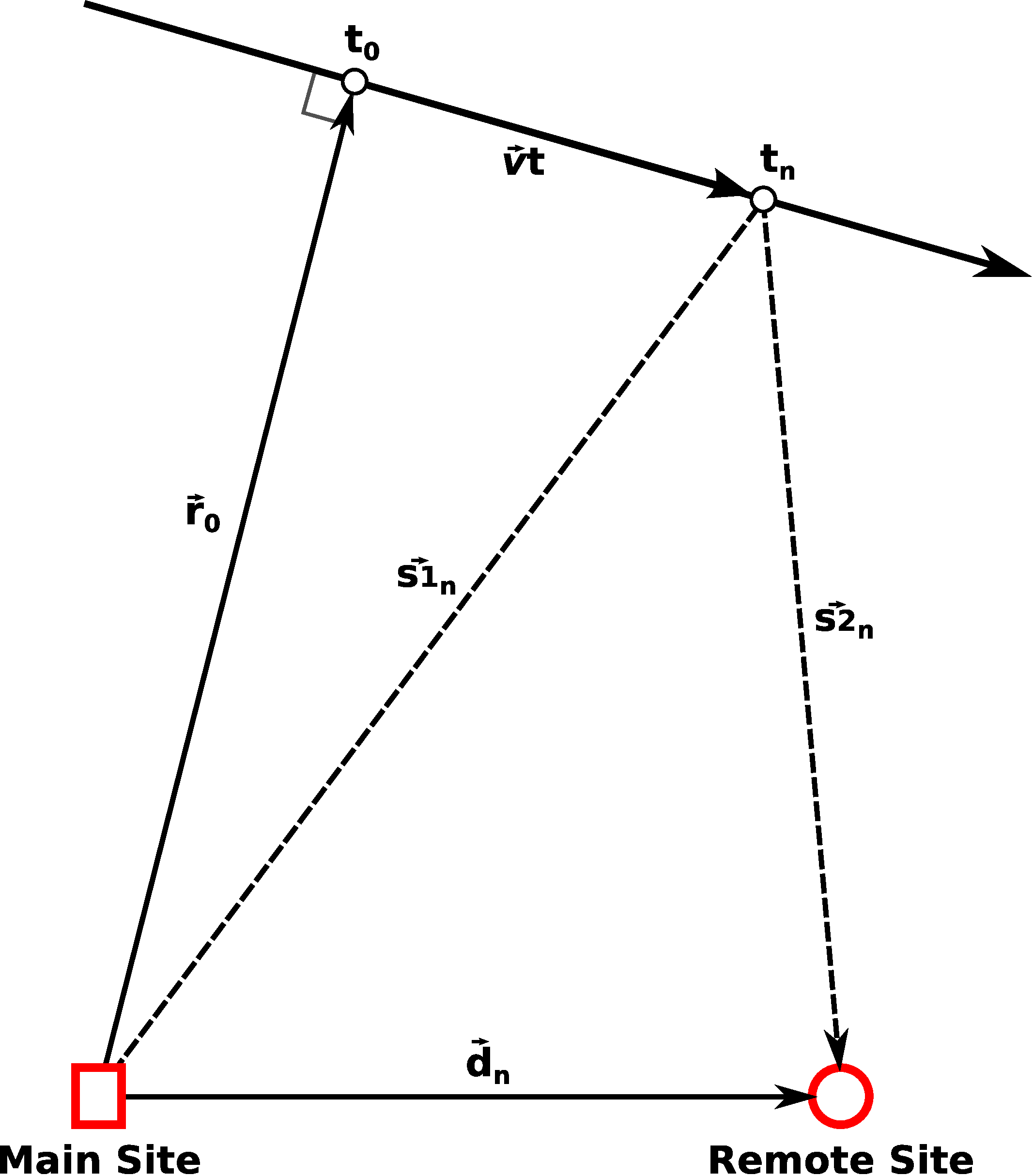}
    \caption{Meteor echo scatter geometry showing radio propagation path from the transmitter (Main Site) to the specular point (R$_{o}$), distance along trail, s, and the $t_0$ / $t_n$ points}
    \label{fig:TOFgeom}
\end{figure}

Figure \ref{fig:TOFgeom} shows the basic geometry for \change[R2]{an}{a} transverse scattering n-station velocity solution. While the main radar site sees a specular echo from the \change[mazur]{t$_0$}{$t_0$} point at range $\Vec{r}_0$, a remote station will see a specular echo from the \change[mazur]{t$_n$}{$t_n$} point at ranges $\Vec{S1}_n$ from the main site and $\Vec{S2}_n$ from the remote site. If follows, then, that the total distance travelled by the radar wave is $R=\mid\Vec{r}_0\mid$ for back-scatter and $S=\mid\Vec{S1}_n\mid+\mid\Vec{S2}_n\mid$ for forward-scatter cases. To solve for the velocity vector, $\Vec{v}$, we need to find the relationship between $\Vec{v}$, $\Vec{d}_n$, and \change[R2]{t}{$t$} that minimizes \change[R2]{S}{$S$}. Since $\Vec{d}_n$ and \change[R2]{t}{$t$}, the transmitter-receiver distance vector and the time between the \change[R2]{t$_n$}{$t_n$} and \change[R2]{t$_0$}{$t_0$} points respectively, are known, the problem can be solved with at least three remote stations. If the problem is further constrained with knowledge of $\Vec{r}_0$, then only two stations are required.

If we consider the general case, S can be expanded \change[mazur]{to }{as},
%%%%%
\begin{linenomath*}
\begin{equation}
\begin{aligned}
    S &= \left\lVert \vec{r_0} + t\vec{v} \right\rVert + \| \vec{r_0} + t\vec{v} + \vec{d} \| \\
    &= [(\vec{r_0} + t\vec{v}) \cdot (\vec{r_0} + t\vec{v})]^{1/2} + [(\vec{r_0} + t\vec{v} - \vec{d}) \cdot (\vec{r_0} + t\vec{v} - \vec{d})]^{1/2}\\
    &= (r_0^2 + 2t\Vec{r}_0 \cdot \Vec{v} + t^2v^2)^{1/2} + (r_0^2 + t^2v^2 + d^2 + 2t\Vec{r}_0 \cdot \Vec{v} - 2t\Vec{v} \cdot \Vec{d})^{1/2}
\label{eq:Sexpanded}
\end{aligned}
\end{equation}
\end{linenomath*}
%%%%%
\add[mazur]{Since we want to minimize $S$ with respect to $t$, we set}\remove[mazur]{Setting} the first time derivative of (\ref{eq:Sexpanded}) to 0 and \change[mazur]{recognizing}{recognize} that $\Vec{r}_0 \cdot \Vec{v}=0$\add[mazur]{. This} gives the following,
%%%%%
\begin{linenomath*}
\begin{equation}
\begin{aligned}
    \frac{dS}{dt} &= (r_0^2 + 2t\vec{r_0} \cdot \vec{v} + t^2v^2)^{-1/2}\\
    &+ (r_0^2 + t^2v^2 + d^2 + 2t\vec{r_0} \cdot \vec{v} - 2\vec{r_0} \cdot \vec{d} - 2t\vec{v} \cdot \vec{d})^{-1/2} (tv^2 + \vec{r_0} \cdot \vec{v} - \vec{v} \cdot \vec{d}) = 0
\label{eq:dSdt}
\end{aligned}
\end{equation}
\end{linenomath*}
%%%%%
which becomes,
%%%%%
\begin{linenomath*}
\begin{equation}
\begin{aligned}
    \| \vec{r_0} + t\vec{v} \| ^{-1}(\vec{r_0} \cdot \vec{v} + tv^2) + \| \vec{r_0} + t\vec{v} - \vec{d} \|^{-1} (tv^2 + \vec{r_0} \cdot \vec{v} - \vec{v} \cdot \vec{d}) &= 0\\
    \|\Vec{r}_0 + t\Vec{v} - \Vec{d}\|tv^2 + \|\Vec{r}_0 + t\Vec{v}\|(tv^2 - \Vec{v} \cdot \Vec{d}) &= 0\\
    \| \vec{r_0} + t\vec{v} - \vec{d} \| (\vec{r_0} \cdot \vec{v} + tv^2) + \| \vec{r_0} + t\vec{v} \| (tv^2 +\vec{r_0} \cdot \vec{v} - \vec{v} \cdot \vec{d}) &= 0
\label{eq:dSdt_2}
\end{aligned}
\end{equation}
\end{linenomath*}
%%%%%
and, since $\vec{r_0}$ is perpendicular to $\vec{v}$, $\vec{r_0} \cdot \vec{v} = 0$ which simplifies (\ref{eq:dSdt_2}),
%%%%%
\begin{linenomath*}
\begin{equation}
\begin{aligned}
    \| \vec{r_0} + t\vec{v} - \vec{d} \|  tv^2 + \| \vec{r_0} + t\vec{v} \| (tv^2 - \vec{v} \cdot \vec{d}) &= 0
\label{eq:dSsimplified}
\end{aligned}
\end{equation}
\end{linenomath*}
%%%%%
\change[mazur]{Using}{For receivers at offsets less than about 20 km, we make} the approximation $\|\Vec{r}_0 + t\Vec{v} - \Vec{d}\| \approx \|\Vec{r}_0 + t\Vec{v}\|$ and \change[mazur]{rearranging}{rearrange} (\ref{eq:dSsimplified}) \change[mazur]{gives}{to give},

%%%%%
\begin{linenomath*}
\begin{equation}
\begin{aligned}
    tv^2 + (tv^2 - \vec{v} \cdot \vec{d}) &= 0\\
    \frac{\vec{v} \cdot \vec{d}}{2v^2} &= t
\label{eq:tsimplified1}
\end{aligned}    
\end{equation}
\end{linenomath*}
%%%%%
By eliminating the range, $\vec{r_0}$, we have also removed interferometry as a requirement for a solution. If we now define $q$ as,
%%%%%
\begin{linenomath*}
\begin{equation}
    \Vec{q} = \frac{\Vec{v}}{2v^2}
\label{eq:q1}
\end{equation}
\end{linenomath*}
%%%%%
we can can simplify (\ref{eq:tsimplified1}) to,
%%%%%
\begin{linenomath*}
\begin{equation}
    \Vec{q} \cdot \Vec{d} = t
\end{equation}
\end{linenomath*}
%%%%%

And since $\Vec{d}$ and \change[mazur]{t}{$t$} are known for each of the \change[R2]{n}{$n$} remote sites, we have a system of \change[R2]{n}{$n$} equations,
%%%%%
\begin{linenomath*}
\begin{equation}
\begin{split}
    \Vec{d}_1 \cdot \Vec{q} & = t_1\\
    \Vec{d}_2 \cdot \Vec{q} & = t_2\\
    \vdots &\\
    \Vec{d}_n \cdot \Vec{q} & = t_n
\end{split}
\end{equation}
\end{linenomath*}
%%%%%
With $\vec{q}$, we can then find $\vec{v}$ by the following,
%%%%%
\begin{linenomath*}
\begin{equation}
\begin{aligned}
    \vec{q} &= \frac{\vec{v}}{2v^2}\\
    \| \vec{q} \| &= \frac{\| \vec{v} \|}{2 \| \vec{v} \| ^2}\\
    \| \vec{q} \| &= \frac{1}{2 \| \vec{v} \|}\\
    \| \vec{v} \| &= \frac{1}{2 \| \vec{q} \|}\\
    v^2 &= \frac{1}{4q^2}\\
    2v^2 &= \frac{1}{2q^2}\\
    \frac{\vec{v}}{\vec{q}} &= \frac{1}{2q^2}\\
    \vec{v} &= \frac{\vec{q}}{2q^2}
\end{aligned}
\label{eq:vfromq}
\end{equation}
\end{linenomath*}
%%%%%
In Cartesian coordinates, $\Vec{d}$ and $\Vec{q}$ have x, y, and z components so the system can be written in matrix form as,
%%%%%
\begin{linenomath*}
\begin{equation}
\begin{pmatrix}
    d_{1x} \;\;\; d_{1y} \;\;\; d_{1z}\\
    d_{2x} \;\;\; d_{2y} \;\;\; d_{2z}\\
    \vdots\\
    d_{nx} \;\;\; d_{2y} \;\;\; d_{2z}
\end{pmatrix}
\begin{pmatrix}
    q_x\\
    q_y\\
    q_z
\end{pmatrix}
    =
\begin{pmatrix}
    t_1\\
    t_2\\
    \vdots\\
    %t_\change[R2]{3}{n}
    t_n %PP: \change is causing errors
\end{pmatrix}
\end{equation}
\end{linenomath*}
%%%%%
Defining the nx3 matrix as $A$ and the column vector $t_1, t_2,\dots,t_n$ at $\Vec{t}$ gives,
%%%%%
\begin{linenomath*}
\begin{equation}
    A = 
\begin{pmatrix}
    d_{1x} \;\;\; d_{1y} \;\;\; d_{1z}\\
    d_{2x} \;\;\; d_{2y} \;\;\; d_{2z}\\
    \vdots\\
    d_{nx} \;\;\; d_{ny} \;\;\; d_{nz}
\end{pmatrix}
    \quad \textrm{and} \quad \Vec{t} =
\begin{pmatrix}
    t_1\\
    t_2\\
    \vdots\\
    t_n
\end{pmatrix}
\label{eq:Aandt_nointerferometry}
\end{equation}
\end{linenomath*}
%%%%%
\begin{linenomath*}
\begin{equation}
    A\Vec{q}=\Vec{t}
\label{eq:Aqt}
\end{equation}
\end{linenomath*}
%%%%%
Multiplying both sides by the transpose of $A$, allows (\ref{eq:Aqt}) to be rearranged so that,
%%%%%
\begin{linenomath*}
\begin{equation}
    \Vec{q}=(A^tA)^{-1}A^t\Vec{t}
\label{eq:q2}
\end{equation}
\end{linenomath*}
%%%%%
Solving (\ref{eq:q2}) with (\ref{eq:Aandt_nointerferometry}), then, allows for the solution for TOF velocity vector for the case with no range information. CMOR, however, uses interferometry at the main station which allows the problem to be further constrained to two dimensions \add[mazur]{where $v$ is forced onto a 2-D plane}.\remove[R2]{ Since $\Vec{r}_0$ is known and is orthogonal to $\Vec{v}$, $\Vec{r}_0 \cdot \Vec{v} = 0$.} And since $\Vec{q}$ is parallel to $\Vec{r}$, it follows that $\Vec{r}_0 \cdot \Vec{q} = 0$. Writing this in Cartesian coordinates and rearranging for $q_z$ gives,
%%%%%
\begin{linenomath*}
\begin{equation}
    q_z=- \bigg( \frac{r_{0x}}{r_{0z}} q_x + \frac{r_{0y}}{r_{0z}} q_y \bigg)
\label{eq:qz}
\end{equation}
\end{linenomath*}
%%%%%
Now, since $q_z$ depends on ratios of the components of $\Vec{r}_0$, the range is not explicitly needed. Instead, zenith angle, $\Theta$, and azimuth, $\Phi$, can be used and (\ref{eq:qz}) can be expressed in terms of $\Theta$ and $\Phi$. The x, y, and z components of $\Vec{r}_0$ are,
%%%%%
\begin{linenomath*}
\begin{equation}
\begin{split}
    r_{0x} = \|\Vec{r}_0\| \sin(\Theta) \cos(\Phi)\\
    r_{0y} = \|\Vec{r}_0\| \sin(\Theta) \sin(\Phi)\\
    r_{0z} = \|\Vec{r}_0\| \cos(\Theta)
\end{split}
\end{equation}
\end{linenomath*}
%%%%%
Which, when substituted into (\ref{eq:qz}) gives,
%%%%%
\begin{linenomath*}
\begin{equation}
    q_z = -(\tan(\Theta)\cos(\Phi)q_x + \tan(\Theta)\sin(\Phi)q_y)
\label{eq:qz2}
\end{equation}
\end{linenomath*}
%%%%%
In Cartesian coordinates the system of equations becomes,
%%%%%
\begin{linenomath*}
\begin{equation}
\begin{aligned}
    q_z &= -(\tan(\Theta)\cos(\Phi)q_x + \tan(\Theta)\sin(\Phi)q_y)\\
    d_{1x}q_x + d_{1y}q_y + d_{1z}q_z &= t_1\\
    d_{2x}q_x + d_{2y}q_y + d_{2z}q_z &= t_2\\
    \vdots\\
    d_{nx}q_x + d_{ny}q_y + d_{nz}q_z &= t_n
    \end{aligned}
\end{equation}
\end{linenomath*}
%%%%%
Which, after eliminating $q_z$ and collecting terms gives,
%%%%%
\begin{linenomath*}
\begin{equation}
\begin{pmatrix}
    d_{1x}-\tan(\Theta)\cos(\Phi)d_{1z}\quad d_{1y}-\tan(\Theta)\sin(\Phi)d_{1z}\\
    d_{2x}-\tan(\Theta)\cos(\Phi)d_{2z}\quad d_{2y}-\tan(\Theta)\sin(\Phi)d_{2z}\\
    \vdots\\
    d_{nx}-\tan(\Theta)\cos(\Phi)d_{nz}\quad d_{ny}-\tan(\Theta)\sin(\Phi)d_{nz}
\end{pmatrix}
\begin{pmatrix}
    q_x\\
    q_y
\end{pmatrix}
    =
\begin{pmatrix}
    t_1\\
    t_2\\
    \vdots\\
    t_3
\end{pmatrix}
\end{equation}
\end{linenomath*}
%%%%%
\add[mazur]{If we then let $A$, $\vec{t}$, and $\vec{q}$ be given as,}
%%%%%
\begin{linenomath*}
\begin{equation}
    A = 
\begin{pmatrix}
    d_{1x} - \tan(\Theta)\cos(\Phi)d_{1z} \;\;\; d_{1y} - \tan(\Theta)\cos(\Phi)d_{1z}\\
    d_{2x} - \tan(\Theta)\cos(\Phi)d_{2z} \;\;\; d_{2y} - \tan(\Theta)\cos(\Phi)d_{2z}\\
    \vdots\\
    d_{nx} - \tan(\Theta)\cos(\Phi)d_{nz} \;\;\; d_{2y} - \tan(\Theta)\cos(\Phi)d_{nz}
\end{pmatrix}
    \quad \textrm{and} \quad \Vec{t} =
\begin{pmatrix}
    t_1\\
    t_2\\
    \vdots\\
    t_n
\end{pmatrix}
    \quad \textrm{and} \quad \Vec{q} =
\begin{pmatrix}
    q_x\\
    q_y\\
\end{pmatrix}
\label{eq:Aandtandq_2Dinterferometry}
\end{equation}
\end{linenomath*}
%%%%%
\add[mazur]{we can write,}
%%%%%
\begin{linenomath*}
\begin{equation}
    A\Vec{q} = \Vec{t}
\end{equation}
\end{linenomath*}
%%%%%
Solving this, along with (\ref{eq:qz2}), gives $q_x$, $q_y$, and $q_z$ without an explicit requirement on range.

To solve in 3 dimensions, range information is added to give the following set of equations.
%%%%%
\begin{linenomath*}
\begin{equation}
\begin{aligned}
    r_{0x}q_x + r_{0y}q_y + r_{0z}q_z &= 0\\
    d_{1x}q_x + d_{1y}q_y + d_{1z}q_z &= t_1\\
    d_{2x}q_x + d_{2y}q_y + d_{2z}q_z &= t_2\\
    \vdots\\
    d_{nx}q_x + d_{ny}q_y + d_{nz}q_z &= t_n
    \end{aligned}
\end{equation}
\end{linenomath*}
%%%%%
Which can be written in matrix form as,
%%%%%
\begin{linenomath*}
\begin{equation}
\begin{pmatrix}
    r_{0x} \;\;\; r_{0y} \;\;\ r_{0z}\\
    d_{1x} \;\;\ d_{1y} \;\;\ d_{1z}\\
    d_{2x} \;\;\ d_{2y} \;\;\ d_{2z}\\
    \vdots\\
    d_{nx} \;\;\ d_{2y} \;\;\ d_{2z}
\end{pmatrix}
\begin{pmatrix}
    q_x\\
    q_y\\
    q_z
\end{pmatrix}
    =
\begin{pmatrix}
    0\\
    t_1\\
    t_2\\
    \vdots\\
    t_3
\end{pmatrix}
\end{equation}
\end{linenomath*}
%%%%%
where,
%%%%%
\begin{linenomath*}
\begin{equation}
    A = 
\begin{pmatrix}
    r_{0x} \;\;\ r_{0y} \;\;\ r_{0z}\\
    d_{1x} \;\;\ d_{1y} \;\;\ d_{1z}\\
    d_{2x} \;\;\ d_{2y} \;\;\ d_{2z}\\
    \vdots\\
    d_{nx} \;\;\ d_{2y} \;\;\ d_{2z}
\end{pmatrix}
    \quad \textrm{and} \quad \Vec{t} =
\begin{pmatrix}
    0\\
    t_1\\
    t_2\\
    \vdots\\
    t_n
\end{pmatrix}
\label{eq:Aandt_3Dinterferometry}
\end{equation}
\end{linenomath*}
%%%%%
\change[mazur]{Again, we find that,}{so that,}
%%%%%
\begin{linenomath*}
\begin{equation}
    A\Vec{q} = \Vec{t}
\end{equation}
\end{linenomath*}
%%%%%
\change[mazur]{So}{From this, we see that} there are three possible TOF solutions which depend on the information available. All cases have the same matrix equation, $A\vec{q}=\vec{t}$, but different expressions for $A$ and $\vec{t}$. These are: 3D with no interferometry (Equation \ref{eq:Aandt_nointerferometry}), 2D interferometry where $v$ is forced to a 2D plane (Equation \ref{eq:Aandtandq_2Dinterferometry}), and 3D interferometry (Equation \ref{eq:Aandt_3Dinterferometry}). \change[mazur]{Solving}{Each of the solutions has given} \remove[mazur]{for} $\vec{q}$ \change[mazur]{then allows for velocity determination}{from which we can determine the meteor's velocity}, $\vec{v}$, with (\ref{eq:vfromq}).

The accuracy of the TOF solution is based on the amount and quality of the input data. For the 3-D solution with no interferometry, $A^TA$ needs to be invertible. This means that $A$ must have a full rank (3 for the 3D case) - a condition which is not met if all sites lie on the same plane. In this case, the 3-D, non-interferometry problem will be poorly defined and will be sensitive to small errors in $t_n$ picks when the sites are close to lying on the same plane. For CMOR, this problem is avoided by using the interferometry, but not range, from the main site to give a 2D solution with interferometry, where $\vec{v}$ is forced onto a 2D plane. In theory, additional precision could be achieved by adding range information. At present, however, range precision is only 1.5 km and would require interpolation within range bins to improve overall precision of perhaps a few percent.
%%%%%
\subsection{CMOR Inflection pick algorithm}
\label{sec:infl}
For CMOR, possible echoes are selected if 15 pulses exceed the noise background by a factor of 2 in any given range gate. From these initial detections, a set of filtering criteria is used to select good quality echoes. These include filters based on: length of rise-time, Fresnel oscillations, duplication of echoes, signal saturation, closeness to the end of the record, multiple peaks, and non-constant interferometry after the peak\add[R1]{ amplitude}. \add[mazur]{This last filter - non-constant interferometry - simply refers to the fact that the interferometry solution is expected to be constant from each window for a given echo. If the solution jumps by more than 3 degrees, multiple specular scattering centres (e.g. from trail distortion by upper atmospheric winds) may be indicated and the echo is rejected.} After filtering, the \add[mazur]{amplitude} inflection point (\remove[mazur]{ideally }equal to the \change[mazur]{t_0}{point of maximum phase} \remove[mazur]{point}) is found on these good echoes by scanning the filtered echo for the point where the second derivative of the amplitude approaches zero. After finding the inflection points for at least three stations, Equation \ref{eq:Aandtandq_2Dinterferometry} is used to solve for $\vec{q}$ and, finally, for $\vec{v}$.

\section{Event Detection}
Events are detected with a program called ``rawproc'', which processes the receiver data streamed to disk, after each 1800 second acquisition cycle has completed. This data is grouped into one second ``chunks'' consisting of all the raw in-phase and quadrature data for each receiver channel at each range gate, as 16-bit signed integers. This software was developed for the study of \citeA{Weryk_Brown_2012}, and works using a three stage process.

First, rawproc loads the positions of each receiver antenna (used by the interferometry), and calculates low-pass filter coefficients (where the corner low-pass frequency is usually 20 Hz) necessary to smooth the amplitude-time data (used by time inflection time pick algorithm). The DC offset level for both the in-phase and quadrature components is recomputed from the first data block by using the median of one second of the respective signal data at a high-numbered range gate. We use high range gates as aircraft echoes are common contamination signals at low range gates.

Next, the threshold values to trigger a detection above the background level are estimated by computing the average and standard deviation of the average amplitude in 12-sample windows over ten range gates at high range. Both the thresholding and DC offset measurements are computed once per streamed file. While these could be computed periodically, they do not vary much during the course of a single acquisition cycle. This method was validated to work through manual inspection of many hours worth of streamed data files verifying each manually identified meteor was found. 

The second stage then locates candidate echoes based on how much \change[R1]{they}{their amplitudes} exceed the background threshold limit. CMOR is configured with a PRF of 532 Hz, and each range gate for each data block is segmented into equal length 14-sample windows. Noise despiking is optionally performed (not necessary for the noise environment of CMOR), the signal data for the current window is incoherently integrated across all five receivers and all range gate windows exceeding the background \add[mazur]{signal} by 8.0 \remove[mazur]{sigma}$\sigma$ are flagged, and a ``count down'' timer of ten windows (140 pulses) set. If a given range gate still exceeds the background level during its next 14-sample window, the timer is reset to ten windows. Once the timer decreases to zero, the candidate event is considered to have ended, and is queued for analysis by the third stage. We note that while 8.0 sigma may seem overly high, the thresholding is based on the variation in average window amplitude. We also note that because CMOR uses wide pulses (to maintain compatibility with the historical three station velocity capability), a candidate event may also be detected in neighbouring range gates. For these cases, only the range gate having the largest amplitude (ie: closest to the centre of the transmitted pulse) is saved, and triggers occurring within two ranges of this maximum are ignored.

The third (and final) stage performs the actual measurements, which may reject candidate events. The software can optionally determine a precise range through interpolation of the range gate data, using a non-linear expression representing the pulse shape used by \citeA{Weryk_Brown_2012}. Next, one second of data is pre-padded, and two seconds appended to the event record. If an event exceeds 3000 samples (5.6 seconds), it is discarded. While this may exclude rare longer duration overdense echoes from being recorded by the system, it much more commonly prevents \change[mazur]{spurious}{non-specular and/or non-meteor} events from being included in the echo data. 

The amplitude vs time data \add[mazur]{for the individual channels} is coherently integrated \add[R2]{in time}, copied to another memory buffer, and filtered using a 255 point 20 Hz low-pass filter. The amplitude inflection point (i.e. the sample corresponding to when the ablating meteoroid reached its maximum phase point) is determined using the algorithm summarised in Appendix \ref{sec:infl}. While some kind of correlation might be better for \change[mazur]{a}{finding the relative} time \change[mazur]{pick}{offsets from the $t_0$ point}, the signal profiles at each remote station occur at different times/heights on the meteor ionization curve, and thus can be morphologically different. As well, \add[mazur]{during initial processing }each station is processed independently \change[mazur]{at which point}{so} the signal data from other stations is not available. The interferometry algorithm (to locate the meteor echo point in local coordinates) is run, and the zenith angle combined with the echo range to give a preliminary height (not relative to the WGS84 geoid).

While many interferometry algorithms are available \cite<e.g.>[]{Jones1998}, they all operate on the same basic principle: the phase offsets between antenna pairs are used to estimate the direction of echo arrival. Some algorithms better handle non co-planar antenna arrangements, and some handle (not applicable to the CMOR setup) \change[R2]{non-consistently}{irregularly} spaced antennas and give uncertainties. 

The baseline interferometry algorithm used by CMOR takes pair-wise solutions across all antennas to estimate the best echo direction similar to the approach described by \citeA{Holdsworth_etal_2004}. The interferometry has been compared to simultaneously detected optical two station detections \cite<e.g.>[]{Weryk_Brown_2012} and found to be accurate to \change[R1]{of order one degree}{~$\sim$1 degree}. % PP: I added the \sim symbol, ~ is just a non-breaking space in LaTeX

The interferometry window size is quite large, as it includes the signal padding. Wind distorted echoes may be affected by this, but these non-classically defined echoes are usually filtered out for astronomical studies. \add[mazur]{In order to calculate the electron line density, the amplitude and power calibration is used to give a power in dBm. }The last steps, then, estimate the background noise power (in dBm), calculate the signal-to-noise ratio of the peak amplitude point, and measure the decay constant and thus estimate the ambipolar diffusion coefficient\remove[mazur]{,} used to correct the peak signal power.\add[R2]{ }This use of the decay constant in this step is necessary to correct the amplitude back to what it would have theoretically been at the $t_0$ point. In this way, the calibrated receiver power will give the true electron line density of the trail. A unique record number is generated which consists of the year, the solar longitude, and a counter number. The measured quantities are written to a log file, and the in-phase and quadrature signal for the range gate closest to the peak are written to a binary record file for subsequent confirmation/analysis.

\section{Event Filtering}

The final stage in accepting a possible echo for \change[R2]{orbit determination}{TOF measurement} involves a more refined set of filter checks.

The echoes found and processed by rawproc are filtered to remove events which are not ideal for accurate velocity measurements. The program ``mevfilt'' reads the output log file from rawproc, and for each echo, loads the binary record file, optionally recomputes the interferometry (using an improved algorithm developed since the data was acquired) and inflection time picks (representing the sample corresponding to when the meteoroid reached its minimum range), and then runs the tests listed in Table \ref{tab:mevfilt}. The tests are performed in order, and the first test to fail will immediately reject an echo. The output log file is identical to that produced by rawproc.

\begin{table}[t]
\caption{Filter tests performed on detected meteor echoes$^{a}$}
\label{tab:mevfilt}
\centering
\begin{tabular}{c c}
\hline
test & description \\
\hline
PASS    & passes all tests\\
NEAREND & peak near ends of record\\
STR     & signal too weak\\
RT      & rise time too long\\
OSC     & signal oscillates after peak\\
PP      & peak after rawproc peak\\
SAT     & signal saturated \\
DUP     & duplicate echo \\
INTERF  & interferometry not constant after peak \\
\hline
\multicolumn{2}{l}{$^{a}$The tests are performed in order, and the first to fail will reject the echo.}
\end{tabular}
\end{table}

%%TC:ignore
\bibliography{Papers}
%%TC:endignore

\end{document}